\documentclass[a4paper,11pt]{article}
\usepackage{jheppub} 

\pdfoutput=1

\usepackage{mathrsfs,amsmath,amsthm,amssymb,xcolor,epstopdf,verbatim,hyperref,enumerate,graphicx,hyperref,subfigure,bm,bbm,amsfonts,subdepth,cleveref,setspace,booktabs,tabularx,placeins,multirow,mathtools,units}
\usepackage{cleveref}
\usepackage[english]{babel}
\usepackage[sort&compress]{natbib}
\usepackage[normalem]{ulem}
\usepackage{dcolumn}
\graphicspath{{Figs/}}
\usepackage{float}
\usepackage{xcolor}
\usepackage{tabularx}

\newcommand{\drawsquare}[2]{\hbox{%
		\rule{#2pt}{#1pt}\hskip-#2pt
		\rule{#1pt}{#2pt}\hskip-#1pt
		\rule[#1pt]{#1pt}{#2pt}}\rule[#1pt]{#2pt}{#2pt}\hskip-#2pt
	\rule{#2pt}{#1pt}}

\newcommand{\fund}{\raisebox{-.5pt}{\drawsquare{6.5}{0.4}}}
\newcommand{\Ysymm}{\raisebox{-.5pt}{\drawsquare{6.5}{0.4}}\hskip-0.4pt%
	\raisebox{-.5pt}{\drawsquare{6.5}{0.4}}}
\newcommand{\Yasymm}{\raisebox{-3.5pt}{\drawsquare{6.5}{0.4}}\hskip-6.9pt%
	\raisebox{3pt}{\drawsquare{6.5}{0.4}}}
\newcommand{\antifund}{\overline{\fund}}

 \renewcommand{\thetable}{\arabic{table}} 
 \def\ov{\overline}
\usepackage{jheppub} 
\def\yzero{\smash{\hbox{$y\kern-4pt\raise1pt\hbox{${}^\circ$}$}}}

\def\ov{\overline}
\def\s2{\frac{1}{\sqrt2}}

\def\beq{\begin{equation}}
	\def\eeq{\end{equation}}
\def\beqa{\begin{eqnarray}}
	\def\eeqa{\end{eqnarray}}

\DeclareMathOperator{\grSU}{SU}
\DeclareMathOperator{\grU}{U}
\DeclareMathOperator{\grUSp}{USp}

\newcommand{\Zbb}{\mathbb{Z}}
\newcommand{\Tbb}{\mathbb{T}}

\begin{document}
	
\title{Generalized Three-Family Supersymmetric Pati-Salam Models from Type IIA Intersecting D6-Branes}

\author[a,1]{Tianjun Li}
\author[b, c, 2]{Qi Sun}
\author[d,3]{Rui Sun}
\author[e,4]{Lina Wu}

\affiliation[a]{School of Physics, Henan Normal University, Xinxiang 453007, P. R. China}

\affiliation[b]{CAS Key Laboratory of Theoretical Physics, Institute of Theoretical Physics,\\
		Chinese Academy of Sciences, Beijing 100190, P. R. China}
\affiliation[c]{School of Physical Sciences, University of Chinese Academy of Sciences,\\
		No.19A Yuquan Road, Beijing 100049, P. R. China}
\affiliation[d]{School of Mathematical Sciences,  University of Chinese Academy of Sciences,\\
		No.19A Yuquan Road, Beijing 100049, P. R. China}
\affiliation[e]{School of Sciences, Xi'an Technological University, Xi'an 710021, P. R. China}

\emailAdd{tli@itp.ac.cn}
\emailAdd{sunqi@itp.ac.cn}
\emailAdd{sunrui24@ucas.ac.cn}
\emailAdd{wulina@xatu.edu.cn}

\abstract{

Generalizing three-family chiral fermion conditions to $I_{ac}=-(3+h)$ and $I_{ac'}=h$, with positive integer $h$, we extend the landscape of  
three-family ${\cal N}=1$ supersymmetric Pati-Salam models in a broader region. Differing from the former investigation with $I_{ac}=-3$ and $I_{ac'}=0$, we do not restrict that the $a$ stack of D6-branes must be parallel to the orientifold image of the $c$-stack along one of the three two-tori. In this investigation, without the simple parallel construction, we find four new classes of supersymmetric Pati-Salam models that are allowed by the extended three generation condition with $I_{ac}=3, I_{ac'}=-6$ and $I_{ac}=-1, I_{ac'}=-2$ through the intersections of $a$- and $c/c'$-branes. 
Moreover, with  the $\grSU(2)_{L'}$ gauge coupling realized from $\grSU(2)_{L_1}\times\grSU(2)_{L_2}$ symmetry breaking, the canonical normalization requirement of the gauge kinetic term provides an  alternative approach that can be imposed before the renormalization group equation evolution for $\grSU(2)_{L'}$ gauge coupling. This turns out to be an effective mechanism to realize the string-scale gauge coupling relation, especially for the new supersymmetric Pati-Salam models with large $g_b/g_a$ ratio. We show that this symmetry-breaking modified renormalization group evolution can highly suppress $g_b/g_a$, and finally realizes string-scale gauge coupling relations for the extended supersymmetric Pati-Salam models as well. 

}

\maketitle

\section{Introduction}

As one of the most important topics in string phenomenology, symmetry breaking and brane intersection theory have been investigated with great efforts from both string model building and phenomenology aspects, such as in~\cite{Bachas:1995ik, Blumenhagen:2000wh, Angelantonj:2000hi,  Aldazabal:2000sa,Aldazabal:2000cn,Aldazabal:2000dg,Ibanez:2001nd,Cvetic:2001tj,Blumenhagen:2002gw}. 
From type IIA  string theory, ${\cal N} = 1$ supersymmetric Pati-Salam models were constructed  with orientifolds $\Tbb^6/(\mathbb{Z}_2\times \mathbb{Z}_2)$ from intersecting D6-branes 
as in~\cite{Berkooz:1996km,Cvetic:1999hb, Cvetic:2000st, Cvetic:2004ui,  Chen:2005aba,Blumenhagen:2005mu, Chen:2007ms}. 
In this construction, the gauge symmetries of ${\cal N} = 1$ supersymmetric Pati-Salam models, $\grSU(4)_C\times \grSU(2)_L \times \grSU(2)_R$, are realized from three stacks of D6-brane intersection, \emph{i.e,} $a$-, $b$-, and $c$-stacks which correspond to $\grSU(4)_C$, $ \grSU(2)_L $, and $\grSU(2)_R$ gauge groups, respectively.
The gauge symmetry can then be broken down to $\grSU(3)_C\times \grSU(2)_L \times \grU(1)_{B-L} \times \grU(1)_{I_{3R}} $ via D6-brane splitting. 
Without introducing any additional anomaly-free $\grU(1)$  around the electroweak scale, this gauge symmetry can be further broken down to Standard Model\,(SM) gauge symmetry via Higgs mechanism. And Yukawa couplings can be generated as well~\cite{Lykken:1998ec, Honecker:2012jd, Honecker:2017air, Anastasopoulos:2006da,Anastasopoulos:2009mr,Anastasopoulos:2010ca, Anastasopoulos:2010hu,Anastasopoulos:2011zz, Ecker:2015vea, Marchesano:2022qbx, Antoniadis:2004dt, Antoniadis:2010zze, Kiritsis:2007zz, Kiritsis:2003mc}. 
In Orbifold and Gepner configurations, D-brane vacua were also studied to realize SM~\cite{Brunner:2004zd}.
From the phenomenology aspects, 
a cornerstone prediction of Grand Unified Theories (GUTs), is naturally realized in the Minimal Supersymmetric Standard Model (MSSM) \cite{Ellis:1990wk,Langacker:1991an,Amaldi:1991cn}, with the unification scale $M_{\text{GUT}} \sim 2 \times 10^{16}$ GeV.
This scale differs from the typical string scale in weakly coupled heterotic string theory, which is an order of magnitude larger. As derived in~\cite{Dienes:1996du}, the string scale is given by
\begin{eqnarray}
    M_{\text{string}}=g_{\text{string}}\times 5.27 \times 10 ^{17} ~\text{GeV},
\end{eqnarray}
where string coupling constant $g_{string}\sim \mathcal{O}(1)$, implying $M_{\text{string}}\simeq 5\times 10^{17}$ GeV. This introduces a factor of $\sim 25$ between $M_{\text{U}}$ and $M_{\text{string}}$, posing a critical challenge in string phenomenology: how to obtain string-scale gauge coupling unification. 

Since ${\cal N} = 1$ supersymmetric Pati-Salam models were constructed from intersecting D6-branes in Type IIA string theory, searching for the complete list of  corresponding  phenomenology models has been a long term topic for string phenomenology. From string theory perspective,  this is related to the important string landscape problem investigated from various perspectives~\cite{Chen:2007px,Chen:2007zu, Cvetic:2003xs,Cvetic:2002pj,Cvetic:2002qa,Cvetic:2002wh,Cvetic:2003yd,Cvetic:2003ch,Honecker:2003vq,Chen:2005mj}.  This has been investigated with random scanning methods, machine learning~\cite{Douglas:2006xy,Halverson:2019tkf, Loges:2021hvn, Loges:2022mao, Cvetic:2004nk, Cvetic:2001nr, Li:2019nvi, Li:2021pxo, Li:2019miw,Li:2022fzt}, and finally via systematic methods by means of solving Diophantine equations for the wrapping numbers, the landscape of the standard Pati-Salam models were completed with 202752 number of ${\cal N}=1$ supersymmetric Pati-Salam models found~\cite{He:2021gug}.
The string-scale coupling relations can also be realized  in~\cite{He:2021kbj, Li:2022cqk} in the following work via two-loop Renormalization Group Equations\,(RGEs) evolution~\cite{Barger:2005qy, Jiang:2006hf, Jiang:2008xrg, Chen:2017rpn,Chen:2018ucf}.
In which, there are in total 33  types of gauge coupling relations presented among the 202752 number of ${\cal N}=1$ supersymmetric Pati-Salam models. In each class, these models are symmetrically related via type I/II T-dualities, D6-brane Sign Equivalent Principle\,(DSEP), and in particular there are four types of equivalent three generation relations due to symmetry transformation, such as $b\leftrightarrow b'$ and $c\leftrightarrow c'$. Therefore, models with quantum numbers $({\bf 4, 2, 1})$ and $({\bf {\bar 4}, 1, 2})$ under the gauge symmetries $\grSU(4)_C\times \grSU(2)_L\times \grSU(2)_R$ 
are considered with three generation conditions constrained by
\begin{eqnarray}
	\label{E3RF} 
    I_{ab} + I_{ab'}~=~3~,~\,\\ \nonumber
 I_{ac} ~=~-3~,~ I_{ac'} ~=~0~, 
\end{eqnarray}
and 
\begin{equation}
\label{E3LF} I_{ac}=0~,~I_{ac'}=-3.
\end{equation}
Note that while $I_{ac'} =0$, the stack $a$ D6-branes must be parallel to the orientifold image of the $c$-stack along one of three two-tori. Similar parallel construction appears for the case of $I_{ac}=0$ as well.

However, although the former landscape search of the standard Pati-Salam models are completed with all the string theory symmetries considered, the above three generation conditions may be further generalized. As proposed in~\cite{Cvetic:2004ui}, in addition to the completed search,  potentially interesting constructions for Pati-Salam models, such as 
\begin{eqnarray}
I_{ac}=-(3+h)~,~ I_{ac'}=h~,~\,
\end{eqnarray}
with positive integer $h$, could also lead to the SM construction of three-family of chiral fermions.
Namely, the former three generation conditions \eqref{E3RF} and \eqref{E3LF} may be simply generalized to 
\begin{eqnarray}\label{gen}
I_{ac}+I_{ac'}=-3~,~\,
\end{eqnarray}
with positive integer $h$, as the generalized three generation conditions.
In this construction, the massless vector-like Higgs fields are allowed, and the  Pati-Salam gauge symmetry can be broken down to
the SM gauge symmetry via the D6-brane splitting and Higgs mechanism.
But it was also pointed in the former investigation that as large wrapping numbers are required to increase the value of $I_{ac}$ and/or $I_{ac'}$, these type of models shall be rare. We refer to~\cite{Cvetic:2004ui} for more detailed computing reason to the conclusion.
That is why in the former investigations these models are not searched broadly in general. Be aware of this omission, here in this paper, we in particular search for the omitted models that allowed by the generalized three generation conditions~\eqref{gen}.

In this new investigation, we also note that there is an isomorphism between the $\grSU(2)$ and $\grUSp(2)$ groups. Such that, the Pati-Salam symmetry $\grSU(4)\times\grSU(2)_L\times\grSU(2)_R$ can be obtained from symmetry breaking of 
$\grSU(4)\times\grSU(2)_L\times\grSU(2)_R\times\grUSp(2)$. 
In which, a diagonal $\grSU(2)_{L'}$ gauge group can be realized from the symmetry breaking of $\grSU(2)_L\times\grUSp(2)$. As a rewarding outcome from this, the ratio of $g_a$ and $g_b$ denoted by $k_2=g_b^2/g_a^2$ can be highly suppressed which helps to realize string-scale gauge coupling unification at $M_{\text{string}}$ from the symmetry-breaking modified renormalization group evolution. In the manner of generic gauge coupling relations, the gauge coupling of the generalized supersymmetric Pati-Salam models can be represented by 
\begin{equation}
	g_a^2=k_2g_b^2=k_Yg_Y^2=g_{\text{U}}^2 \sim g_{\text{string}}^2,
\end{equation}
where $k_Y$ and $k_2$ are model-dependent constants that determine the weak mixing angle $\sin \theta_W$ at the string scale. 

The paper is organized as follows. In Section 2,  we
briefly review the basic rules for supersymmetric model building from intersecting D6-branes on Type IIA orientifolds, with 
the tadpole cancellation conditions and $N=1$ supersymmetry condition presented. 
In section 3, we introduce the generalization three-family conditions and the extended landscape of the supersymmetric Pati-Salam models. The $\grSU(2)_L\times\grUSp(2)$ symmetry breaking which helps to suppress the gauge coupling is proposed in the end of this section as well.
In section 4, we study in detail of the string-scale gauge coupling relation for the new Pati-Salam models with Renormalization Group Equations (RGEs) evolution.
Finally, Section 5 is devoted to the conclusion and outlook, with the T-dual of the new Pati-Salam models presented in the Appendix.

\section{Intersecting D6-branes Construction on $\Tbb^6 /(\Zbb_2 \times \Zbb_2)$ Orientifolds}

D6-branes intersecting at generic angles on Type IIA $\Tbb^6 /(\Zbb_2 \times \Zbb_2)$ orientifolds have been investigated intensively to construct supersymmetric Pati-Salam models, such as in~\cite{Cvetic:2002pj, Cvetic:2001nr}.
Here the six-torus $\Tbb^{6}$ can be decomposed into  three two-tori  $\Tbb^{6} = \Tbb^{2} \times \Tbb^{2} \times \Tbb^{2}$, and thus 
the complex coordinates in the two-dimensional torus can be denoted as $z_i$, $i=1,\; 2,\; 3$, respectively.
The generators of the orbifold group $\Zbb_{2} \times \Zbb_{2}$, $\theta$ and $\omega$, correspond to the two vectors $(1/2,-1/2,0)$ and $(0,1/2,-1/2)$, respectively. They act on the complex coordinates $z_i$, such that 
\beqa
& \theta: & (z_1,z_2,z_3) \mapsto (-z_1,-z_2,z_3)~,~ \nonumber \\
& \omega: & (z_1,z_2,z_3) \mapsto (z_1,-z_2,-z_3)~.~\,
\label{orbifold} \eeqa 
The orientifold projection is defined by gauging the $\Omega R$ symmetry, where $\Omega$ denotes the world-sheet parity and $R$ represents the map with
\beqa
R: (z_1,z_2,z_3) \mapsto ({\ov z}_1,{\ov z}_2,{\ov
	z}_3)~.~\, 
\eeqa
The projection actions  $\Omega R$, $\Omega R\theta$, $\Omega R\omega$, and $\Omega R\theta\omega$ lead to different kinds of orientifold 6-planes (O6-planes).
Moreover, $N_a$  stacks of D6-branes wrapping on the factorized three-cycles can be introduced  to cancel the RR charges of the O6-planes.
And there are two possible complex structures that are consistent with orientifold projection on each two-torus:
rectangular or tilted~\cite{Cvetic:2001nr, Chen:2007zu, Cvetic:2002pj, Blumenhagen:2000ea}. 
The homology classes are represented by $n_a^i[a_i]+m_a^i[b_i]$ and $n_a^i[a'_i]+m_a^i[b_i]$ which correspond to rectangular and tilted tori, respectively, with $[a_i']=[a_i]+\frac{1}{2}[b_i]$.
Thus, in both cases, two wrapping numbers $(n_a^i,l_a^i)$ are used to label a generic one cycle,
where $l_{a}^{i}\equiv m_{a}^{i}$ for a rectangular two-torus and $l_{a}^{i}\equiv 2\tilde{m}_{a}^{i}=2m_{a}^{i}+n_{a}^{i}$ on a tilted two-torus. 
Therefore, for tilted two-tours it is necessary that $l_a^i-n_a^i$ must be even.

In addition, for $a$-stack of $N_a$ D6-branes along the cycle $(n_a^i,l_a^i)$, their $\Omega R$ images as $a'$-stack of $N_a$ D6-branes are then labeled with wrapping numbers $(n_a^i,-l_a^i)$.
And their homology three-cycles are represented by
\beq
[\Pi_a]=\prod_{i=1}^{3}\left(n_{a}^{i}[a_i]+2^{-\beta_i}l_{a}^{i}[b_i]\right)\quad \text{and} \quad
\left[\Pi_{a'}\right]=\prod_{i=1}^{3}
\left(n_{a}^{i}[a_i]-2^{-\beta_i}l_{a}^{i}[b_i]\right)~,~\, \eeq
where $\beta_i=0$ when the $i$-th torus is rectangular and $\beta_i=1$ for tilted torus.
The homology three-cycles wrapping on the four  kinds of O6-planes can be represented by
\beq \Omega R: [\Pi_{\Omega R}]= 2^3
[a_1]\times[a_2]\times[a_3]~,~\, \eeq \beq \Omega R\omega:
[\Pi_{\Omega
	R\omega}]=-2^{3-\beta_2-\beta_3}[a_1]\times[b_2]\times[b_3]~,~\,
\eeq \beq \Omega R\theta\omega: [\Pi_{\Omega
	R\theta\omega}]=-2^{3-\beta_1-\beta_3}[b_1]\times[a_2]\times[b_3]~,~\,
\eeq \beq
\Omega R\theta:  [\Pi_{\Omega
	R}]=-2^{3-\beta_1-\beta_2}[b_1]\times[b_2]\times[a_3]~.~\,
\label{orienticycles} \eeq 
Consequently, the intersection numbers between different stacks of D6-branes can be represented in terms of wrapping numbers such that 
\beq \label{ab}
I_{ab}=[\Pi_a][\Pi_b]=2^{-k}\prod_{i=1}^3(n_a^il_b^i-n_b^il_a^i)~,~\,
\eeq \beq
I_{ab'}=[\Pi_a]\left[\Pi_{b'}\right]=-2^{-k}\prod_{i=1}^3(n_{a}^il_b^i+n_b^il_a^i)~,~\,
\eeq \beq
I_{aa'}=[\Pi_a]\left[\Pi_{a'}\right]=-2^{3-k}\prod_{i=1}^3(n_a^il_a^i)~,~\,
\eeq \beq {I_{aO6}=[\Pi_a][\Pi_{O6}]=2^{3-k}(-l_a^1l_a^2l_a^3
	+l_a^1n_a^2n_a^3+n_a^1l_a^2n_a^3+n_a^1n_a^2l_a^3)}~,~\,
\label{intersections} \eeq

where $k=\beta_1+\beta_2+\beta_3$ denotes the total number of tilted two-tori while $\beta_i=1$ for tilted tori and $\beta_i=0$ for rectangular tori.
And $[\Pi_{O6}]=[\Pi_{\Omega R}]+[\Pi_{\Omega R\omega}]+[\Pi_{\Omega
R\theta\omega}]+[\Pi_{\Omega R\theta}]$ represents the sum of homology three-cycles associated with four O6-planes.

The generic massless particle spectrum on the intersecting D6-branes at general angles can be represented by the intersection numbers as summarized in Table~\ref{spectrum}, which holds for both rectangular and tilted two-tori configurations.  The second column in the table specifies the representations under the gauge group $\grU(N_a/2)$ surviving under the $\Zbb_2\times \Zbb_2$ orbifold projection~\cite{Cvetic:2001nr}.  
All chiral supermultiplets contain both scalar and fermionic components in the supersymmetric configurations, and positive intersection numbers are adopted to denote the left-handed chiral supermultiplets.

\begin{table}[h] 
	\renewcommand{\arraystretch}{1.25}
	\begin{center}
		\begin{tabular}{|c|c|}
			\hline {\bf Sector} & \phantom{more space inside this box}{\bf
				Representation}
			\phantom{more space inside this box} \\
			\hline\hline
			$aa$   & $\grU(N_a/2)$ vector multiplet  \\
			& 3 adjoint chiral multiplets  \\
			\hline
			$ab+ba$   & $I_{ab}$ $(\fund_a,\antifund_b)$ fermions   \\
			\hline
			$ab'+b'a$ & $I_{ab'}$ $(\fund_a,\fund_b)$ fermions \\
			\hline $aa'+a'a$ &$\frac 12 (I_{aa'} - \frac 12 I_{a,O6})\;\;
			\Ysymm\;\;$ fermions \\
			& $\frac 12 (I_{aa'} + \frac 12 I_{a,O6}) \;\;
			\Yasymm\;\;$ fermions \\
			\hline
		\end{tabular}
	\end{center}
	\caption{ 
	The general massless particle spectrum arising from D6-branes intersecting at generic angles is presented. 
}
	\label{spectrum}
\end{table}

\subsubsection*{The four-dimensional ${\cal N}=1$ Supersymmetric Conditions}

In four-dimensional ${\cal N}=1$ supersymmetric models, $1/4$ supercharges from ten-dimensional Type I T-dual are restricted to be preserved. This preservation is required under the orientation projection of the intersecting D6-branes and the $\Zbb_2\times \Zbb_2$ orbifold projection on the background manifold.
Here, any D6-brane rotation angle with respect to the orientifold plane shall be an element of $\grSU(3)$ for four-dimensional ${\cal N}=1$ supersymmetry to survive under the orientation projection as in~\cite{Berkooz:1996km}. 
This is equivalent to $\theta_1+\theta_2+\theta_3=0 $ mod $2\pi$, where $\theta_i$ is the angle between the $D6$-brane and the orientifold-plane in the $i$-th two-torus.

Moreover, we denote the products of wrapping numbers in a compact way to simplify the following discussion, such as 
\beq
\begin{array}{rrrr}
	A_a \equiv -n_a^1n_a^2n_a^3, & B_a \equiv n_a^1l_a^2l_a^3,
	& C_a \equiv l_a^1n_a^2l_a^3, & D_a \equiv l_a^1l_a^2n_a^3, \\
	\tilde{A}_a \equiv -l_a^1l_a^2l_a^3, & \tilde{B}_a \equiv
	l_a^1n_a^2n_a^3, & \tilde{C}_a \equiv n_a^1l_a^2n_a^3, &
	\tilde{D}_a \equiv n_a^1n_a^2l_a^3.\,
\end{array}
\label{variables}\eeq 
Due to the $\Zbb_2\times \Zbb_2$ orbifold projection automatically survives under such D6-brane configuration, the four-dimensional ${\cal N}=1$ supersymmetry conditions read as in~\cite{Cvetic:2002pj}
\begin{eqnarray}
	x_A\tilde{A}_a+x_B\tilde{B}_a+x_C\tilde{C}_a+x_D\tilde{D}_a=0,
	\nonumber\\\nonumber \\ A_a/x_A+B_a/x_B+C_a/x_C+D_a/x_D<0,
	\label{susyconditions}
\end{eqnarray} 
where $x_A=\lambda,\;
x_B=\lambda 2^{\beta_2+\beta3}/\chi_2\chi_3,\; x_C=\lambda
2^{\beta_1+\beta3}/\chi_1\chi_3,\; x_D=\lambda
2^{\beta_1+\beta2}/\chi_1\chi_2$, 
and the complex structure moduli with the $i$-th two-torus can be written in terms of $\chi_i=R^2_i/R^1_i$.  
And a positive parameter $\lambda$ can be introduced to put all variables $A,\,B,\,C,\,D$ in an equivalent position.

\subsubsection*{The RR Tadpole Cancellation Conditions}

As one of the most important conditions for intersecting D-brane construction, the tadpole cancellation conditions directly lead to the $\grSU(N_a)^3$ cubic non-Abelian anomaly cancellation.
And as discussed in~\cite{Cvetic:2001nr,Aldazabal:2000dg,Ibanez:2001nd},
the Green-Schwarz mechanism mediated by untwisted RR fields can be used to cancel the $\grU(1)$ mixed gauge and gravitational anomalies or $[\grSU(N_a)]^2 \grU(1)$ gauge anomalies.

The RR fields arise from D6-branes and orientifold O6-planes which are restricted by the Gauss law in a compact space.
The RR charges of D6-branes and D6-planes shall be canceled for conservation of the RR field flux lines, and thus
the RR tadpole cancellation conditions can be represented by
\begin{eqnarray}
	\sum_a N_a [\Pi_a]+\sum_a N_a
	\left[\Pi_{a'}\right]-4[\Pi_{O6}]=0~,~\,
\end{eqnarray}
where the last terms are due to the O6-planes with $-4$ RR charges in the D6-brane charge unit. 
To cancel the RR tadpoles, an arbitrary number of D6-branes called "filler branes" wrapping along the orientifold planes can be introduced.
And the RR tadpole cancellation conditions can be rewritten as 
\begin{eqnarray}
	\label{eq:tadpole}
	-2^k N^{(1)}+\sum_a N_a A_a=-2^k N^{(2)}+\sum_a N_a
	B_a= \nonumber\\ -2^k N^{(3)}+\sum_a N_a C_a=-2^k N^{(4)}+\sum_a
	N_a D_a=-16,\,
\end{eqnarray}
where $2 N^{(i)}$ corresponds to the number of filler branes wrapping the $i$-th O6-plane. 
In addition, the filler branes realize the $\grUSp$ gauge group and share identical wrapping numbers with one of the four O6-planes as shown in Table~\ref{orientifold1}.
They also trivially contribute to the four-dimensional ${\cal N}=1$ supersymmetry conditions.
\begin{table}[h] 
	\begin{center}
		\begin{tabular}{|c|c|c|}
			\hline
			Orientifold Action & O6-Plane & $(n^1,l^1)\times (n^2,l^2)\times
			(n^3,l^3)$\\
			\hline
			$\Omega R$& 1 & $(2^{\beta_1},0)\times (2^{\beta_2},0)\times
			(2^{\beta_3},0)$ \\
			\hline
			$\Omega R\omega$& 2& $(2^{\beta_1},0)\times (0,-2^{\beta_2})\times
			(0,2^{\beta_3})$ \\
			\hline
			$\Omega R\theta\omega$& 3 & $(0,-2^{\beta_1})\times
			(2^{\beta_2},0)\times
			(0,2^{\beta_3})$ \\
			\hline
			$\Omega R\theta$& 4 & $(0,-2^{\beta_1})\times (0,2^{\beta_2})\times
			(2^{\beta_3},0)$ \\
			\hline
		\end{tabular}
	\end{center}
	\caption{The wrapping numbers for four O6-planes.} \vspace{0.4cm}
	\label{orientifold1}
\end{table}

Based on the above conditions, all possible D6-brane configurations can be classified into three categories under four-dimensional ${\cal N}=1$ supersymmetry condition. The $\grUSp$ groups can be divided into $A$-, $B$-, $C$- or $D$-types based on whether their associated filler branes carry non-zero $A$, $B$, $C$ or $D$, respectively.  

(1) In the case that the filler brane has the same wrapping numbers as one of the O6-planes, the gauge symmetry is $\grUSp$ group and only one wrapping number product $A$, $B$, $C$ and $D$ has non-zero and negative value.
We denote the corresponding $\grUSp$ group as the $A$-, $B$-, $C$- or $D$-type $\grUSp$ group according to which one is non-zero.

(2) In the case that there is a zero wrapping number, it is a Z-type D6-brane.
$A$, $B$, $C$ and $D$ have two positive and two negative.

(3) In the case that there is no zero wrapping number, it is a NZ-type D6-brane. 
$A$, $B$, $C$ and $D$ have one positive and three negative.
We denote NZ-type branes in the $A$-, $B$-, $C$- and $D$-type NZ branes by the positive one.
Each type can be further divided into two subtypes by the wrapping numbers taking the form as follows
\begin{eqnarray}
	A1: (-,-)\times(+,+)\times(+,+),~& A2:(-,+)\times(-,+)\times(-,+);\\
	B1: (+,-)\times(+,+)\times(+,+),~& B2:(+,+)\times(-,+)\times(-,+);\\
	C1: (+,+)\times(+,-)\times(+,+),~& C2:(-,+)\times(+,+)\times(-,+);\\
	D1: (+,+)\times(+,+)\times(+,-),~& D2:(-,+)\times(-,+)\times(+,+).
\end{eqnarray}
For convenience, we call Z-type and NZ-type D6-branes as $U$-branes because they have gauge symmetry $\grU(n)$.

\section{Generalization of three-family Supersymmetric Pati-Salam Models}

Two extra $\grU(1)$ gauge groups are needed when one obtains SM or standard-like models by intersecting D6-branes. 
As a result, right-handed charged leptons obtain the correct quantum number in both supersymmetric and non-supersymmetric models~\cite{Ibanez:2001nd,Cvetic:2001nr,Cvetic:2003xs,Cvetic:2002pj}.
These two groups are, respectively, the lepton number symmetry $\grU(1)_L$ and the third component of the right-handed weak isospin  $\grU(1)_{I_{3R}}$. And the hypercharge $Q_Y$ is related 
 to the the baryonic charge $Q_B$ as
\begin{eqnarray}
Q_Y = Q_{I_{3R}} + \frac{Q_B - Q_L}{2} ,
\end{eqnarray}
where $Q_B$ is generated by the decomposition $\grU(3)_C \simeq \grSU(3)_C \times \grU(1)_B$.
The gauge group $\grU(1)_{I_{3R}}$ comes from the non-abelian component of the symmetry $\grU(2)_R$ or $\grUSp$ due to the massless gauge field of $\grU(1)_{I_{3R}}$.
Otherwise, $\grU(1)_{I_{3R}}$ will get mass from the $B \wedge F$ couplings.
Similarly, an anomaly-free $\grU(1)_{B-L}$ must come from the non-abelian group, and in previous related work on supersymmetric model building, $\grU(1)_{I_{3R}}$ comes from the $\grUSp$ groups~\cite{Cvetic:2001nr,Cvetic:2003xs}.

These models actually possess two anomaly-free $\grU(1)$ symmetries and are configured with no less than eight Higgs doublets. 
Although it is theoretically possible to achieve the breaking of its symmetry group to SM, this process will inevitably lead to violating of the D-flatness and F-flatness, thereby inevitably losing supersymmetry.
We consider a configuration of three distinct D6-brane stacks denoted by a, b, and c, comprising 8, 4, and 4 D6-branes, respectively. 
The associated gauge symmetries are $\grU(4)_C \times \grU(2)_L$ and $\grU(2)_R$, giving rise to the Pati–Salam gauge group $\grSU(4)_C \times \grSU(2)_L \times \grSU(2)_R$. 
As outlined in~\cite{Cvetic:2004ui}, through a combination of D6-brane splitting and the Higgs mechanism, this symmetry can be broken down to the SM gauge group according to the following symmetry-breaking chain:

\begin{eqnarray}
\grSU(4)\times \grSU(2)_L \times \grSU(2)_R  &&
\xrightarrow{\;a\rightarrow a_1+a_2\;}\;  \grSU(3)_C\times \grSU(2)_L
\times \grSU(2)_R \times \grU(1)_{B-L} \nonumber\\&&
\xrightarrow{\; c\rightarrow c_1+c_2 \;} \; \grSU(3)_C\times \grSU(2)_L\times
\grU(1)_{I_{3R}}\times \grU(1)_{B-L} \nonumber\\&&
\xrightarrow{\;\text{Higgs Mechanism}\;}\; \grSU(3)_C\times \grSU(2)_L\times \grU(1)_Y~.~\,
\end{eqnarray}
The phenomenon of dynamical supersymmetry breaking has been investigated in~\cite{Cvetic:2003yd} within the framework of D6-brane constructions derived from Type IIA orientifolds. 
For a D6-brane stack labeled a, the corresponding kinetic function takes the following form~\cite{Chen:2007zu}:
\begin{equation}
f_a = \frac{1}{4\kappa_a}\left(n_{a}^{1}n_{a}^{2}n_{a}^{3}s - \frac{n_{a}^{1}l_{a}^{2}l_{a}^{3}u^{1}}{2^{\beta_{2}+\beta_{3}}} - \frac{l_{a}^{1}n_{a}^{2}l_{a}^{3}u^{2}}{2^{\beta_{1}+\beta_{3}}} - \frac{l_{a}^{1}l_{a}^{2}n_{a}^{3}u^{3}}{2^{\beta_{1}+\beta_{2}}}\right),
\end{equation}
where $\kappa_a$ is a constant with respect to the gauge groups, as an example $\kappa_a$ = 1 for $\grSU(N_a)$.
We work with the moduli parameter $s$ and $u^i$ (with $i = 1,2,3$) in the supergravity basis, which are connected to the four-dimensional dilaton $\phi_4$ and the moduli parameters of complex structure $U_i$ (with $i = 1,2,3$) through the following relations:
\begin{equation}
\begin{aligned}
\operatorname{Re}(s) &= \frac{e^{-\phi_{4}}}{2\pi}\frac{\sqrt{\operatorname{Im}(U^{1})\operatorname{Im}(U^{2})\operatorname{Im}(U^{3})}}{|U^{1}U^{2}U^{3}|}, \\
\operatorname{Re}(u^{1}) &= \frac{e^{-\phi_{4}}}{2\pi}\sqrt{\frac{\operatorname{Im}(U^{1})}{\operatorname{Im}(U^{2})\operatorname{Im}(U^{3})}}\left|\frac{U^{2}U^{3}}{U^{1}}\right|, \\
\operatorname{Re}(u^{2}) &= \frac{e^{-\phi_{4}}}{2\pi}\sqrt{\frac{\operatorname{Im}(U^{2})}{\operatorname{Im}(U^{1})\operatorname{Im}(U^{3})}}\left|\frac{U^{1}U^{3}}{U^{2}}\right|, \\
\operatorname{Re}(u^{3}) &= \frac{e^{-\phi_{4}}}{2\pi}\sqrt{\frac{\operatorname{Im}(U^{3})}{\operatorname{Im}(U^{1})\operatorname{Im}(U^{2})}}\left|\frac{U^{1}U^{2}}{U^{3}}\right|.
\end{aligned}
\end{equation}
In the current model, the complex structures $U^i$ (with $i = 1,2,3$) can be written in terms of the moduli parameters as~\cite{Chen:2007zu}
\begin{equation}
U^1 = i\chi_1, ~U^2 = i\chi_2, ~ U^3 = \frac{2\chi_3^2 + 4i\chi_3}{4 + \chi_3^2} ,
\end{equation}
where  $\chi_1, \chi_2$, and $\chi_3$ are not independent degrees of freedom, as they can be fully expressed through the parameters $x_A,x_B,x_C,x_D$ which themselves are constrained by the supersymmetric condition given in~\eqref{susyconditions}.
 In practice, these $\chi_i$ parameters are determined only up to an overall coefficient, reflecting freedom under a dilation transformation. 
The Kähler potential is given by
\begin{equation}
K = -\ln(S+\overline{S}) - \sum_{i=1}^{3}\ln(U^{i}+\overline{U}^{i}).
\end{equation}
The complete determination of all moduli parameters, therefore, requires the stabilization of this dilation degree of freedom.

In this approach, gaugino condensation is often utilized to fix the overall coefficient, and at least two $\grUSp$ groups are usually presented in the hidden sector such as discussed in ~\cite{Taylor:1990wr,Brustein:1992nk,deCarlos:1992kox}. And 
the one-loop beta functions are described in~\cite{Cvetic:2004ui} as 
\begin{equation}
\beta_i^g = -3\left(\frac{N^{(i)}}{2} + 1\right) + 2|I_{ai}| + |I_{bi}| + |I_{ci}| + 3\left(\frac{N^{(i)}}{2} - 1\right)
         = -6 + 2|I_{ai}| + |I_{bi}| + |I_{ci}|.
\end{equation}

For each $\grUSp(N^{(i)})$ group originated from $2N^{(i)}$ filler branes, the corresponding one-loop beta functions must be negative.
In the present work, we do not confine our analysis to configurations containing at least two $\grUSp$ groups in the hidden sectors, to accommodate other prospective symmetry-breaking mechanisms.
The gauge coupling constant associated with stack $a$ of the D6-branes can be given by
\begin{equation}
g_{a}^{-2} = |\mathrm{Re}(f_{a})|\, ,
\end{equation}
and as follows the gauge couplings for stacks $b$ and $c$ of the D6-branes are obtained in the same way.
The kinetic function associated with the $\grU(1)_Y$ gauge group is a linear combination of the kinetic functions corresponding to the $\grSU(4)_C$ and $\grSU(2)_R$ groups, as established in~\cite{Blumenhagen:2000ea,Chen:2007zu}
\begin{equation}
f_{Y} = \frac{3}{5}\left(\frac{2}{3}f_{a} + f_{c}\right),
\end{equation}
while the coupling constant $g_Y$ is determined by 
\begin{equation}
g_{Y}^{-2} = |\mathrm{Re}(f_{Y})|.
\end{equation}
At the tree level, the gauge coupling relation reads
\begin{equation}\label{gaugecoupling}
g_{a}^{2} = \alpha g_{b}^{2} = \beta \frac{5}{3}g_{Y}^{2} = \gamma (\pi e^{\phi_{4}}),
\end{equation}
where $\alpha$, $\beta$, and $\gamma$ denote the ratios between the strong, weak, and hypercharge couplings, respectively. And here the $\phi^4$ represents the dilaton field.

\subsection{Intersecting Branes Model Building}

Recall that for brane intersecting construction, a systematic searching algorithm has been constructed in~\cite{He:2021gug}, with the standard Pati-Salam landscape completely given. In which, each Pati-Salam model is determined by 18 integer wrapping numbers  $n_a^1,\dotsc, n_c^3, l_a^1,\dotsc, l_c^3$.

In our new  Pati-Salam model building with the generalized three-family condition $I_{ab}+I_{ab'}=3, I_{ac}+I_{ac'}=-3$, these wrapping numbers are still restricted by the condition of tadpole cancellation~\eqref{eq:tadpole}, supersymmetry condition~\eqref{susyconditions}. 
Still, the wrapping number relation $l_a^1 - n_a^1$ shall be even while the corresponding torus is tilted.
Utilizing the first parts of the systematic searching method in~\cite{He:2021gug}, we revise for the generalized Pati-Salam model building as follows:

First step: we still list the sign of wrapping number products $A_a,B_a,\dotsc,C_c,C_d$.
As discussed in \cite{Cvetic:2002pj, He:2021gug}, due to the supersymmetry constraints, there are three type of possible signs for wrapping number products $(A_a,B_a,C_a,D_a)$ of  each $a/b/c-$ stacks of brane construction: three negative numbers and one positive number;
two negative numbers and two zeroes;
one negative number and three zeros. 

Second step: for each possible sign in the first step, we divide the whole problem into categories according to  the inequalities of wrapping number products, such as $A_a>0,B_a<0,C_a<0,D_a<0$, and try to solve the values of the wrapping numbers.
In each category, we first solve the generalized three family relation such that:
\begin{enumerate}
	\item Equation of the form 
   $I_{ac}+I_{ac'}=-3$. This includes additional 
   $I_{ac} = -1, I_{ac'}=-2$, or $I_{ac}=3, I_{ac'}=-6$,..., (other than the usual $I_{ac} = -3, I_{ac'}=0$),  and vice verse for the value of $I_{ac}, \text{and} I_{ac'}$. From the definition of $I_{ab}$ and $I_{ab'}$, we consider  $\prod_{i=1}^3(n_a^i l_c^i - l_a^i n_c^i)$, $\prod_{i=1}^3(n_a^i l_c^i + l_a^i n_c^i)$ and naturally take $n_a^i l_c^i - l_a^i n_c^i$, $i=1,2,3$ as variables.

	\item Under the additional solutions of above, solve the equation of 	$I_{ab}+I_{ab'}=3$. Again, consider  $\prod_{i=1}^3(n_a^i l_c^i - l_a^i n_c^i)$, $\prod_{i=1}^3(n_a^i l_c^i + l_a^i n_c^i)$ and naturally take $n_a^i l_c^i - l_a^i n_c^i$, $i=1,2,3$ as variables. 
    
	\item Different with the former system of linear equations of full rank, the new solution of model building does not include 
	\[\left\{\begin{array}{c}
		n_a^1 l_c^1- l_a^1 n_c^1=0\\
		n_a^1 l_c^1 + l_a^1 n_c^1=6,
	\end{array}\right.\]
	while
    	\[\left\{\begin{array}{c}
		n_a^1 l_b^1- l_a^1 n_b^1=0\\
		n_a^1 l_b^1 + l_a^1 n_b^1=6,
	\end{array}\right.\]
    are still presented in this extension with $n_a^1 l_b^1$ and $l_a^1 n_b^1$  as variables. From the former research, we know that this kind of equations has unique solutions.

    \item For each type of the linear inequalities derived from the first step, we could still see a subsystem in the form of  
	\[\left\{\begin{array}{r}
		4+2A_a + A_b+A_c \geq 0\\
		A_a<0\\
		A_b<0\\
		A_c=0.
	\end{array}\right.\]
	For each type of linear inequalities listed in Step 1, the integers solution system corresponds to a polyhedron, and whether the polyhedron has finite volume corresponds to whether we have a finite number of solutions/models.
\end{enumerate}
According to the restrictions from these inequalities, the total number of unknown wrapping numbers decrease. Part of the systems are split into subsystems as here the inequalities can not uniquely determine the 
values of the 18 wrapping numbers.

Third step: By repeating the Second step for each type of inequalities, we find 4 classes of new Pati-Salam models as show in Table~\ref{model1}, \ref{model2}, \ref{model3}, \ref{model4} from the three-family generalization as follows\footnote{Here we note that for each model, there is a class of physical equivalent generalized models according to the gauge coupling relations.}. 

\begin{table}[!h]
\scriptsize
	\caption{D6-brane configurations and intersection numbers of Model 1, and its MSSM gauge coupling relation is $g^2_a=\frac{7}{6}g^2_b=\frac{35}{66}g^2_c=\frac{175}{268}(\frac{5}{3}g^2_Y)=\frac{8 \sqrt[4]{2} 5^{3/4} \pi  e^{\phi_4}}{11 \sqrt{3}}$.}
	\label{model1}
	\begin{center}
		\begin{tabular}{|c||c|c||c|c|c|c|c|c|c|}
			\hline\rm{Model} 1 & \multicolumn{9}{c|}{$U(4)\times U(2)_L\times U(2)_R\times USp(2) $}\\
			\hline \hline			\rm{stack} & $N$ & $(n^1,l^1)\times(n^2,l^2)\times(n^3,l^3)$ & $n_{\Ysymm}$& $n_{\Yasymm}$ & $b$ & $b'$ & $c$ & $c'$ & 1\\
			\hline
			$a$ & 8 & $(1,1)\times (1,0)\times (1,-1)$ & 0 & 0  & 3 & 0 & 3 & -6 & 0\\
			$b$ & 4 & $(1,-1)\times (1,-1)\times (1,2)$ & -2 & -6  & - & - & 8 & 0 & 2\\
			$c$ & 4 & $(-1,5)\times (0,1)\times (-1,2)$ & 9 & -9  & - & - & - & - & 10\\
			\hline
			1 & 2 & $(2, 0)\times (1, 0)\times (1, 0)$& \multicolumn{7}{c|}{$x_A = \frac{1}{12} x_B = \frac{1}{10} x_C= \frac{1}{12}x_D$}\\
			& & & \multicolumn{7}{c|}{$\beta^g_1=6$}\\
			& & & \multicolumn{7}{c|}{$\chi_1=\frac{1}{\sqrt{10}}$,  $\chi_2=\frac{\sqrt{\frac{5}{2}}}{6}$, $\chi_3=\sqrt{\frac{2}{5}}$}\\
			\hline
		\end{tabular}
	\end{center}
\end{table}

\begin{table}[!h]\scriptsize
	\caption{D6-brane configurations and intersection numbers of Model 2, and its MSSM gauge coupling relation is $g^2_a=\frac{11}{6}g^2_b=\frac{5}{14}g^2_c=\frac{25}{52}(\frac{5}{3}g^2_Y)=\frac{8}{63} 5^{3/4} \sqrt{11} \pi  e^{\phi_4}$.}
	\label{model2}
	\begin{center}
		\begin{tabular}{|c||c|c||c|c|c|c|c|c|c|c|}
			\hline\rm{Model} 2 & \multicolumn{10}{c|}{$U(4)\times U(2)_L\times U(2)_R\times USp(2)^2 $}\\
			\hline \hline			\rm{stack} & $N$ & $(n^1,l^1)\times(n^2,l^2)\times(n^3,l^3)$ & $n_{\Ysymm}$& $n_{\Yasymm}$ & $b$ & $b'$ & $c$ & $c'$ & 1 & 4\\
			\hline
			$a$ & 8 & $(-1,1)\times (-1,1)\times (-1,1)$ & 0 & 4  & 0 & 3 & 3 & -6 & 1 & 1\\
			$b$ & 4 & $(-2,1)\times (0,1)\times (-1,1)$ & -1 & 1  & - & - & 9 & -10 & 1 & 0\\
			$c$ & 4 & $(-1,2)\times (-1,0)\times (5,1)$ & 9 & -9  & - & - & - & - & 0 & 1\\
			\hline
			1 & 2 & $(1, 0)\times (1, 0)\times (2, 0)$& \multicolumn{8}{c|}{$x_A = 22 x_B = 2 x_C= \frac{11}{5}x_D$}\\
			4 & 2 & $(0, 1)\times (0, 1)\times (2, 0)$& \multicolumn{8}{c|}{$\beta^g_1=-3$, $\beta^g_4=-3$}\\
			& & & \multicolumn{8}{c|}{$\chi_1=\frac{1}{\sqrt{5}}$, $\chi_2=\frac{11}{\sqrt{5}}$, $\chi_3=4 \sqrt{5}$}\\
			\hline
		\end{tabular}
	\end{center}
\end{table}

\begin{table}[!h]\scriptsize
	\caption{D6-brane configurations and intersection numbers of Model 3, and its MSSM gauge coupling relation is $g^2_a=3g^2_b=\frac{13}{5}g^2_c=\frac{65}{41}(\frac{5}{3}g^2_Y)=\frac{16}{5} \sqrt{3} \pi  e^{\phi_4}$.}
	\label{model3}
	\begin{center}
		\begin{tabular}{|c||c|c||c|c|c|c|c|c|c|c|}
			\hline\rm{Model} 3 & \multicolumn{10}{c|}{$U(4)\times U(2)_L\times U(2)_R\times USp(2)\times USp(4) $}\\
			\hline \hline			\rm{stack} & $N$ & $(n^1,l^1)\times(n^2,l^2)\times(n^3,l^3)$ & $n_{\Ysymm}$& $n_{\Yasymm}$ & $b$ & $b'$ & $c$ & $c'$ & 1 & 3\\
			\hline
			$a$ & 8 & $(-1,1)\times (-1,0)\times (1,1)$ & 0 & 0  & 3 & 0 & -1 & -2 & 0 & 0\\
			$b$ & 4 & $(-1,4)\times (0,1)\times (-1,1)$ & 3 & -3  & - & - & -8 & 4 & 4 & 1\\
			$c$ & 4 & $(1,0)\times (1,-1)\times (1,3)$ & -2 & 2  & - & - & - & - & 0 & 1\\
			\hline
			1 & 2 & $(1, 0)\times (1, 0)\times (2, 0)$& \multicolumn{8}{c|}{$x_A = \frac{3}{4} x_B = \frac{1}{4} x_C= \frac{3}{4}x_D$}\\
			3 & 4 & $(0, 1)\times (1, 0)\times (0, 2)$& \multicolumn{8}{c|}{$\beta^g_1=-2$, $\beta^g_3=-4$}\\
			& & & \multicolumn{8}{c|}{$\chi_1=\frac{1}{2}$, $\chi_2=\frac{3}{2}$, $\chi_3=1$}\\
			\hline
		\end{tabular}
	\end{center}
\end{table}

\begin{table}[!h]\scriptsize
	\caption{D6-brane configurations and intersection numbers of Model 4, and its MSSM gauge coupling relation is $g^2_a=6g^2_b=\frac{26}{5}g^2_c=\frac{130}{67}(\frac{5}{3}g^2_Y)=\frac{16}{5} \sqrt{6} \pi  e^{\phi_4}$.}
	\label{model4}
	\begin{center}
		\begin{tabular}{|c||c|c||c|c|c|c|c|c|c|}
			\hline\rm{Model} 4 & \multicolumn{9}{c|}{$U(4)\times U(2)_L\times U(2)_R\times USp(4) $}\\
			\hline \hline			\rm{stack} & $N$ & $(n^1,l^1)\times(n^2,l^2)\times(n^3,l^3)$ & $n_{\Ysymm}$& $n_{\Yasymm}$ & $b$ & $b'$ & $c$ & $c'$ & 3\\
			\hline
			$a$ & 8 & $(-1,1)\times (-1,0)\times (1,1)$ & 0 & 0  & 3 & 0 & -1 & -2 & 0\\
			$b$ & 4 & $(-1,4)\times (0,1)\times (-1,1)$ & 3 & -3  & - & - & -16 & 8 & 1\\
			$c$ & 4 & $(1,0)\times (2,-1)\times (1,3)$ & -5 & 5  & - & - & - & - & 1\\
			\hline
			3 & 4 & $(0, 1)\times (1, 0)\times (0, 2)$& \multicolumn{7}{c|}{$x_A = \frac{3}{2} x_B = \frac{1}{4} x_C= \frac{3}{2}x_D$}\\
			& & & \multicolumn{7}{c|}{$\beta^g_3=-4$}\\
			& & & \multicolumn{7}{c|}{$\chi_1=\frac{1}{2}$, $\chi_2=3$, $\chi_3=1$}\\
			\hline
		\end{tabular}
	\end{center}
\end{table}

\subsection{Spectrum of Generalized Pati-Salam models}

The gauge symmetry for these four classes of extended Pati-Salam models is still $\grU(4)\times \grU(2)_L\times \grU(2)_R$ as the standard construction, yet with only one, and two confining groups. 
Different with the former Pati-Salam models, three and four confining groups are not observed.
While there are four confining gauge groups in the hidden sector with four negative $\beta$ functions appearing, one can break supersymmetry via gaugino condensation. 
For models in Table~\ref{model2} and \ref{model3}, there are two negative $\beta$ functions with two confining $\grUSp(N)$ gauge groups, a general analysis of  the non-perturbative superpotential with tree-level gauge couplings and moduli stabilization according to dilaton and complex structure moduli can be investigated, as discussed in \cite{Cvetic:2003yd}.
However, we shall note that these extrema from moduli stabilization
might be saddle points that do not lead to supersymmetry breaking, while at the stable extrema the supersymmetry can be broken in general.

Now we take the representative models of these 4 classes in Table~\ref{model1} to \ref{model4} as examples to discuss the full spectrum explicitly.  Their gauge coupling unification from RGEs will be left to discuss in the next section with  $\grSU(3)_C \times \grSU(2)_L\times \grU(1)_Y$  gauge coupling unification at the near string scale. 

\begin{table}
[!h] \footnotesize
\renewcommand{\arraystretch}{1.0}
\caption{The chiral spectrum in the open string sector of Model 1} \label{spectrum Model 1}
\begin{center}
\begin{tabular}{|c||c||c|c|c||c|c|c|}\hline
Model 1 & $SU(4)\times SU(2)_L\times SU(2)_R \times USp(2)$
& $Q_4$ & $Q_{2L}$ & $Q_{2R}$ & $Q_{em}$ & $B-L$ & Field \\
\hline\hline
$ab$ & $3 \times (4,\overline{2},1,1)$ & 1 & -$1$ & 0  & $-\frac 13,\; \frac 23,\;-1,\; 0$ & $\frac 13,\;-1$ & $Q_L, L_L$\\
$ac$ & $3 \times (4,1,\overline{2},1)$ & $1$ & 0 & -$1$   & $-\frac 13,\; \frac 23,\; -1,\; 0$ & $\frac 13,\; -1$ & $\overline{Q_R}, \overline{L_R}$\\
$ac'$ & $6 \times (\overline{4},1,\overline{2},1)$ & -$1$ & 0 & -$1$   & $\frac 13,\; -\frac 23,\; 1,\; 0$ & $-\frac 13,\; 1$ & $Q_R, L_R$\\
$bc$ & $8 \times(1,2,\overline{2},1)$ & 0 & $1$ & -$1$   & $1,\;0,\;0,\;-1$ & 0 & $H'$\\
$b1$ & $2\times (1,2,1,\overline{2})$ & $0$ & 1 & 0   & $\pm \frac 12$ & 0 & \\
$c1$ & $10\times(1,1,2,\overline{2})$ & 0 & 0 & 1   & $\pm \frac 12$ & 0 & \\
$b_{\overline{\Ysymm}}$ & $2\times(1,\overline{3},1,1)$ & 0 & -2 & 0   & $0,\pm 1$ & 0 & \\
$b_{\overline{\Yasymm}}$ & $6\times(1,\overline{1},1,1)$ & 0 & -2 & 0   & 0 & 0 & \\
$c_{\Ysymm}$ & $9\times(1,1,{3},1)$ & 0 & 0 & 2   & $0,\pm 1$ & 0 & \\
$c_{\overline{\Yasymm}}$ & $9\times(1,1,\overline{1},1)$ & 0 & 0 & -2   & 0 & 0 & \\
	\hline\hline
$bc'$ & $3 \times (1,2,{2},1)$ & 0 & 1 & 1   & { $1,\;0,\; 0,\;-1$} &0 &$H_u^i, H_d^i$\\
& ${3} \times (1,\overline{2},\overline{2},1)$ & 0 & -1 & -1   &  & &  \\
\hline
\end{tabular}
\end{center}
\end{table}
From the spectrum of Model 1, the Higgs up and down doublets arise from the intersection of $b$-stack and the image of $c$-stack of branes. However, as the $\beta$ function of $\grUSp(2)$ group is positive, it is not confining which leads to difficulty for dynamical supersymmetry breaking in model building. 
Therefore, we do not focus on this class of models for further phenomenology study.
For the spectrum of Model 2, the Higgs doublets cannot arise from the intersection of $b$-stack and $c$-stack of branes, either the intersection of $b$-stack and the image of $c$-stack of branes. However, the $\beta$ function of its two $\grUSp(2)$ groups are negative with possible confining groups leading to stable minimum for supersymmetry breaking. 
\begin{table}
[!h] \footnotesize
\renewcommand{\arraystretch}{1.0}
\caption{The chiral spectrum in the open string sector of Model 2} \label{spectrum Model 2}
\begin{center}
\begin{tabular}{|c||c||c|c|c||c|c|c|}\hline
Model 2 & $SU(4)\times SU(2)_L\times SU(2)_R \times USp(2)^2$
& $Q_4$ & $Q_{2L}$ & $Q_{2R}$ & $Q_{em}$ & $B-L$ & Field \\
\hline \hline
$ab'$ & $3 \times (4,2,1,1,1)$ & 1 & $1$ & 0  & $-\frac 13,\; \frac 23,\;-1,\; 0$ & $\frac 13,\;-1$ & $Q_L, L_L$\\
$ac$ & $3 \times (4,1,\overline{2},1,1)$ & $1$ & 0 & -$1$   & $-\frac 13,\; \frac 23,\;-1,\; 0$ & $\frac 13,\;-1$ & $\overline{Q_R}, \overline{L_R}$\\
$ac'$ & $6 \times (\overline{4},1,\overline{2},1,1)$ & -$1$ & 0 & -$1$   & $\frac 13$,\; $-\frac 23,\; 1,\; 0$ & $-\frac 13,\; 1$ & $Q_R, L_R$\\
$bc$ & $9 \times(1,2,\overline{2},1,1)$ & 0 & $1$ & -$1$   & $1,\;0,\;0,\;-1$ & 0 & $H'$\\
$bc'$ & $10 \times(1,\overline{2},\overline{2},1,1)$ & 0 & -$1$ & -$1$   & $1,\;0,\;0,\;-1$ & 0 & $H''$\\
$a1$ & $1\times (4,1,1,\overline{2},1)$ & $1$ & 0 & 0 & $\frac 16,\;-\frac 12$ & $\frac 13,\;-1$ & \\
$a4$ & $1\times (4,1,1,1,\overline{2})$ & $1$ & 0 & 0   & $\frac 16,\;-\frac 12$ & $\frac 13,\;-1$ & \\
$b1$ & $1\times (1,2,1,\overline{2},1)$ & $0$ & 1 & 0   & $\pm\frac 12$ & 0 & \\
$c4$ & $1\times(1,1,2,1,\overline{2})$ & 0 & 0 & 1   & $\pm \frac 12$ & 0 & \\
$a_{\Yasymm}$ & $4\times(6,1,1,1,1)$ & 2 & 0 & 0   & $-\frac 13, 1$ & $-\frac 23,2$ & \\
$b_{\overline{\Ysymm}}$ & $1\times(1,\overline{3},1,1,1)$ & 0 & -$2$ & 0   & $0,\pm 1$ & 0 & \\
$b_{\Yasymm}$ & $1\times(1,{1},1,1,1)$ & 0 & 2 & 0   & 0 & 0 & \\
$c_{\Ysymm}$ & $9\times(1,1,{3},1,1)$ & 0 & 0 & 2   & $0,\pm 1$ & 0 & \\
$c_{\overline{\Yasymm}}$ & $9\times(1,1,\overline{1},1,1)$ & 0 & 0 & -2   & 0 & 0 & \\
\hline
\end{tabular}
\end{center}
\end{table}

\begin{table}
[!h] \footnotesize
\renewcommand{\arraystretch}{1.0}
\caption{The chiral spectrum in the open string sector of Model 3} \label{spectrum Model 3}
\begin{center}
\begin{tabular}{|c||c||c|c|c||c|c|c|}\hline
Model 3 & $SU(4)\times SU(2)_L\times SU(2)_R $
& $Q_4$ & $Q_{2L}$ & $Q_{2R}$ & $Q_{em}$ & $B-L$ & Field \\
 & $\times USp(2)\times
USp(4)$
& & & & & & \\
\hline\hline
$ab$ & $3 \times (4,\overline{2},1,1,1)$ & 1 & -$1$ & 0  & $-\frac 13,\; \frac 23,\;-1,\; 0$ & $\frac 13,\;-1$ & $Q_L, L_L$\\
$ac$ & $1 \times (\overline{4},1,2,1,1)$ & -$1$ & 0 & $1$   & $\frac 13,\; -\frac 23,\;1,\; 0$ & $-\frac 13,\;1$ & $Q_R, L_R$\\
$ac'$ & $2 \times (\overline{4},1,\overline{2},1,1)$ & -$1$ & 0 & -$1$   & $\frac 13,\; -\frac 23,\;1,\; 0$ & $-\frac 13,\;1$ & $Q_R, L_R$\\
$bc$ & $8 \times(1,\overline{2},2,1,1)$ & 0 & -$1$ & $1$   & $1,\;0,\;0,\;-1$ & 0 & $H'$\\
$bc'$ & $4 \times(1,2,2,1,1)$ & 0 & $1$ & $1$   & $1,\;0,\;0,\;-1$ & 0 & $H$\\
$b1$ & $4\times(1,2,1,\overline{2},1)$ & 0 & 1 & 0   & $\pm \frac 12$ & 0 & \\
$b3$ & $1\times(1,2,1,1,\overline{4})$ & 0 & 1 & 0   & $\pm \frac 12$ & 0 & \\
$c3$ & $1\times(1,1,2,1,\overline{4})$ & 0 & 0 & 1   & $\pm \frac 12$ & 0 & \\
$b_{\Ysymm}$ & $3\times(1,3,1,1,1)$ & 0 & $2$ & 0   & $0,\pm 1$ & 0 & \\
$b_{\overline{\Yasymm}}$ & $3\times(1,\overline{1},1,1,1)$ & 0 & -2 & 0   & 0 & 0 & \\
$c_{\overline{\Ysymm}}$ & $2\times(1,1,\overline{3},1,1)$ & 0 & 0 & -2   & $0,\pm 1$ & 0 & \\
$c_{\Yasymm}$ & $2\times(1,1,1,1,1)$ & 0 & 0 & 2   & 0 & 0 & \\
\hline
\end{tabular}
\end{center}
\end{table}

In the class of Model 3, the representative model has 8 Higgs multiplets naturally arising from the intersection of $b$-stack and $c$-stack of branes, also with relatively few exotic particles beyond SM. 
The $\beta$ functions of its $\grUSp(2)$ and $\grUSp(4)$ groups are negative, with possible confining groups leading to a stable minimum for supersymmetry breaking as well. 

\begin{table}
	[!h] \scriptsize
	\renewcommand{\arraystretch}{1.0}
	\caption{The composite particle spectrum of Model 3, which is		formed due to the strong forces from hidden sector.}
	\label{Composite3}
	\begin{center}
		\begin{tabular}{cc c c}\hline
			\multicolumn{2}{c }{Model 3} &
			\multicolumn{2}{c}{$\grSU(4)\times \grSU(2)_L\times \grSU(2)_R \times
				\grUSp(2)\times
				\grUSp(4)$} \\
			\hline Confining Force & Intersection & Exotic Particle
			Spectrum & Confined Particle Spectrum \\
			\hline
			$\grUSp(2)_1$ &$b1$ & $4\times (1,2,1,2,1)$ & $8\times (1,2^2,1,1,1)$\\
			\hline
			$\grUSp(2)_3$ &$b3$ & $1\times (1,2,1,1,2)$ & $1\times (1,2^2,1,1,1)$, $1\times(1,2,2,1,1)$\\
			&$c3$ & $1\times(1,1,2,1,2)$ &  $1\times(1,1,2^2,1,1)$\\
			\hline
		\end{tabular}
	\end{center}
\end{table}

Model 3 has two confining gauge groups, $\grUSp(2)_1$
and $\grUSp(4)_3$. 
For $\grUSp(2)_2$, there is only one charged intersection,
and $\grUSp(4)_3$  has two charged intersections. Therefore, for $\grUSp(2)_1$, there is no mixed-confinement, their self-confinement leads to 8 tensor representations for each of them. While $\grUSp(4)_3$ has two charged intersections, and therefore besides self-confinement, the mixed-confinement between different intersections also exist, which yields the chiral supermultiplets $(1,2,2,1,1)$. 
Moreover, these spectra of models are anomaly-free as there is no new anomaly introduced to the remaining gauge symmetry~\cite{Cvetic:2002qa}.

\begin{table}[!h]
\footnotesize
\renewcommand{\arraystretch}{1.0}
\caption{The chiral spectrum in the open string sector of Model 4} \label{spectrum Model 4}
\begin{center}
\begin{tabular}{|c||c||c|c|c||c|c|c|}\hline
Model 4 & $SU(4)\times SU(2)_L\times SU(2)_R \times
USp(4)$
& $Q_4$ & $Q_{2L}$ & $Q_{2R}$ & $Q_{em}$ & $B-L$ & Field \\
\hline\hline
$ab$ & $3 \times (4,\overline{2},1,1)$ & 1 & -$1$ & 0  & $-\frac 13,\; \frac 23,\;-1,\; 0$ & $\frac 13,\;-1$ & $Q_L, L_L$\\
$ac$ & $1 \times (\overline{4},1,2,1)$ & -$1$ & 0 & $1$   & $\frac 13,\; -\frac 23,\;1,\; 0$ & $-\frac 13,\;1$ & $Q_R, L_R$\\
$ac'$ & $2 \times (\overline{4},1,\overline{2},1)$ & -$1$ & 0 & -$1$   & $\frac 13,\; -\frac 23,\;1,\; 0$ & $-\frac 13,\;1$ & $Q_R, L_R$\\
$bc$ & $16 \times(1,\overline{2},2,1)$ & 0 & -$1$ & $1$   & $1,\;0,\;0,\;-1$ & 0 & $H'$\\
$bc'$ & $8 \times(1,{2},{2},1)$ & 0 & $1$ & $1$   & $1,\;0,\;0,\;-1$ & 0 & $H$\\
$b3$ & $1\times(1,2,1,\overline{4})$ & 0 & 1 & 0   & $\pm \frac 12$ & 0 & \\
$c3$ & $1\times(1,1,2,\overline{4})$ & 0 & 0 & 1   & $\pm \frac 12$ & 0 & \\
$b_{\Ysymm}$ & $3\times(1,3,1,1)$ & 0 & $2$ & 0   & $0,\pm 1$ & 0 & \\
$b_{\overline{\Yasymm}}$ & $3\times(1,\overline{1},1,1)$ & 0 & -2 & 0   & 0 & 0 & \\
$c_{\overline{\Ysymm}}$ & $5\times(1,1,\overline{3},1)$ & 0 & 0 & -2   & $0,\pm 1$ & 0 & \\
$c_{\Yasymm}$ & $5\times(1,1,1,1)$ & 0 & 0 & 2   & 0 & 0 & \\
\hline
\end{tabular}
\end{center}
\end{table}

\begin{table}
	[!h] \scriptsize
	\renewcommand{\arraystretch}{1.0}
	\caption{The composite particle spectrum of Model 4, which is		formed due to the strong forces from hidden sector.}
	\label{Composite4}
	\begin{center}
		\begin{tabular}{cc c c}\hline
			\multicolumn{2}{c }{Model 4} &
			\multicolumn{2}{c}{$\grSU(4)\times \grSU(2)_L\times \grSU(2)_R \times
				\grUSp(4)$} \\
			\hline Confining Force & Intersection & Exotic Particle
			Spectrum & Confined Particle Spectrum \\
			\hline
			$\grUSp(4)_3$ &$b3$ & $1\times (1,2,1,2)$ & $1\times (1,2^2,1,1)$, $1\times(1,2,2,1)$\\
			&$c3$ & $1\times(1,1,2,2)$ &  $1\times(1,1,2^2,1)$\\
			\hline
		\end{tabular}
	\end{center}
\end{table}
For the class of Model 4, the representative model in Table~\ref{model4} has the spectrum in Table~\ref{spectrum Model 4} with 8 Higgs multiplets naturally arising from the intersection of $b$-stack and the image of $c$-stack of branes. The $\beta$ function of its single $\grUSp(4)$ group is negative with possible confining groups leading to stable minimum for supersymmetry breaking as well.
As $\grUSp(4)_3$ has two charged intersections, and thus besides self-confinement leading to tensor representations, the mixed-confinement between different intersection exists and yields to the chiral supermultiplets $(1,2,2,1)$.

\subsection{Gauge Coupling Relation Modified by Symmetry Breaking}
\label{sec:symm}

The $\grSU(2)_{L_1}\times\grSU(2)_{L_2}$ symmetry can be broken down to the diagonal subgroup $\grSU(2)_{L'}$ through the mechanism of vacuum expectation values.
Specifically, if the gauge couplings of the two subgroups $\grSU(2)_{L_1}$ and $\grSU(2)_{L_2}$ are $g_{L_1}$ and $g_{L_2}$, respectively, the coupling constant $g_{L'}$ of the resulting diagonal subgroup $\grSU(2)$ after breaking satisfies
\begin{equation}
\frac{1}{g_{L'}} = \frac{1}{g_{L_1}} + \frac{1}{g_{L_2}}.
\end{equation}
This relation stems from the canonical normalization requirement of the gauge kinetic terms after symmetry breaking. 
Within the framework of the holomorphic gauge kinetic function, the gauge kinetic function for the diagonal subgroup can be written as
\begin{equation}
f_{L'} = f_{L_1} + f_{L_2}.
\end{equation}
In this way, to some extent, we can suppress the value of the $g_{L'}$ coupling for the overall $\grSU(2)_{L}$.
For instance, we can introduce the filler brane on type-1 O6-plane as the $d$-stack.
Taking Model 2 as an example,
due to the isomorphism between the $\grSU(2)$ and $\grUSp(2)$ groups, we can obtain the Pati-Salam model symmetry $\grSU(4)\times\grSU(2)_L\times\grSU(2)_R$ from $\grSU(4)\times\grSU(2)_L\times\grSU(2)_R\times\grUSp(2)_1$ by the mechanism of symmetry breaking from $\grSU(2)_L\times\grUSp(2)_1$ to the diagonal $\grSU(2)_{L'}$.
By comparing the gauge coupling relations of Model 2 and the suppressed gauge coupling relations resulting from symmetry breaking as shown in Model 2-m in Table~\ref{model2-m}, one can see that the value of the $g_{L'}$ coupling, which is $g_b$ in Model 2, is significantly suppressed.

\begin{table}[!h]\scriptsize
	\caption{D6-brane configurations and intersection numbers of Model 2-m, and its MSSM gauge coupling relation is $g^2_a=\frac{55}{723}g^2_L=\frac{5}{14}g^2_c=\frac{25}{52}(\frac{5}{3}g^2_Y)=\frac{8}{63} 5^{3/4} \sqrt{11} \pi  e^{\phi_4}$.}
	\label{model2-m}
	\begin{center}
		\begin{tabular}{|c||c|c||c|c|c|c|c|c|c|c|}
			\hline\rm{Model} 2-m & \multicolumn{10}{c|}{$U(4)\times U(2)_L\times U(2)_R\times USp(2)_1\times USp(2) $}\\
			\hline \hline			\rm{stack} & $N$ & $(n^1,l^1)\times(n^2,l^2)\times(n^3,l^3)$ & $n_{\Ysymm}$& $n_{\Yasymm}$ & $b$ & $b'$ & $c$ & $c'$ & 1 & 4\\
			\hline
			$a$ & 8 & $(-1,1)\times (-1,1)\times (-1,1)$ & 0 & 4  & 0 & 3 & 3 & -6 & 1 & 1\\
			$b$ & 4 & $(-2,1)\times (0,1)\times (-1,1)$ & -1 & 1  & - & - & 9 & -10 & 1 & 0\\
			$c$ & 4 & $(-1,2)\times (-1,0)\times (5,1)$ & 9 & -9  & - & - & - & - & 0 & 1\\
			\hline
			1 & 2 & $(1, 0)\times (1, 0)\times (2, 0)$& \multicolumn{8}{c|}{$x_A = 22 x_B = 2 x_C= \frac{11}{5}x_D$}\\
			4 & 2 & $(0, 1)\times (0, 1)\times (2, 0)$& \multicolumn{8}{c|}{$\beta^g_1=-3$, $\beta^g_4=-3$}\\
			& & & \multicolumn{8}{c|}{$\chi_1=\frac{1}{\sqrt{5}}$, $\chi_2=\frac{11}{\sqrt{5}}$, $\chi_3=4 \sqrt{5}$}\\
			\hline
		\end{tabular}
	\end{center}
\end{table}

Using the same symmetry breaking method as processed for Model 2, we can obtain better gauge coupling relation closer to unification with $\grSU(2)_L$ gauge coupling suppressed for Model 3 to $\grSU(2)_{L'}$ gauge coupling relation in Model 3-m, as shown in Table~\ref{model3-m}.
Performing symmetry breaking to suppress the high $\grSU(2)_L$ gauge coupling first, will help to achieve gauge couplings unification via RGEs method as shown in the next section.

\begin{table}[!h]\scriptsize
	\caption{D6-brane configurations and intersection numbers of Model 3-m, and its MSSM gauge coupling relation is $g^2_a=\frac{24}{23}g^2_L=\frac{13}{5}g^2_c=\frac{65}{41}(\frac{5}{3}g^2_Y)=\frac{16}{5} \sqrt{3} \pi  e^{\phi_4}$.}
	\label{model3-m}
	\begin{center}
		\begin{tabular}{|c||c|c||c|c|c|c|c|c|c|c|}
			\hline\rm{Model} 3-m & \multicolumn{10}{c|}{$U(4)\times U(2)_L\times U(2)_R\times USp(2)_1\times USp(4) $}\\
			\hline \hline			\rm{stack} & $N$ & $(n^1,l^1)\times(n^2,l^2)\times(n^3,l^3)$ & $n_{\Ysymm}$& $n_{\Yasymm}$ & $b$ & $b'$ & $c$ & $c'$ & 1 & 3\\
			\hline
			$a$ & 8 & $(-1,1)\times (-1,0)\times (1,1)$ & 0 & 0  & 3 & 0 & -1 & -2 & 0 & 0\\
			$b$ & 4 & $(-1,4)\times (0,1)\times (-1,1)$ & 3 & -3  & - & - & -8 & 4 & 4 & 1\\
			$c$ & 4 & $(1,0)\times (1,-1)\times (1,3)$ & -2 & 2  & - & - & - & - & 0 & 1\\
			\hline
			1 & 2 & $(1, 0)\times (1, 0)\times (2, 0)$& \multicolumn{8}{c|}{$x_A = \frac{3}{4} x_B = \frac{1}{4} x_C= \frac{3}{4}x_D$}\\
			3 & 4 & $(0, 1)\times (1, 0)\times (0, 2)$& \multicolumn{8}{c|}{$\beta^g_1=-2$, $\beta^g_3=-4$}\\
			& & & \multicolumn{8}{c|}{$\chi_1=\frac{1}{2}$, $\chi_2=\frac{3}{2}$, $\chi_3=1$}\\
			\hline
		\end{tabular}
	\end{center}
\end{table}

\section{Gauge Unification of Pati-Salam Generalization}

With the above four classes of new supersymmetric Pati-Salam models obtained from generalized three-family of chiral fermion condition(represented by Model 1,2,3,4), and symmetry breaking mechanism introduced, here in this section we study whether the important gauge coupling unification presented in~\cite{He:2021kbj, Li:2022cqk} 
can still be realized.
Before we discuss the string-scale gauge coupling unification in the current generalization, let us first review the basics to obtain gauge unification in supersymmetric Pati-Salam models from the former study. 

In~\cite{He:2021kbj, Li:2022cqk}, the string-scale gauge coupling relations were systematically investigated for all the independent 33 supersymmetric Pati-Salam models from the former searches. With the decoupling of the exotic particles discussed, two-loop RGEs evolution methods are applied to approach string-scale gauge coupling relation by introducing different types of additional particles. 
These particles include additional particles from the adjoint representations of $SU(4)_C$ and $SU(2)_L$ gauge symmetries, SM vector-like particles from four-dimensional chiral sectors, and vector-like particles from $\mathcal{N}=2$ subsector. 
In such a way, the total 33 independent supersymmetric Pati-Salam models all managed to arrive string-scale gauge coupling unification with the above additional particles introduced. Therefore, although these models do not directly have a traditional gauge coupling unification at string scale, an effective approach to realize the string-scale gauge coupling unification was established.

Recall that in the supersymmetric Pati-Salam models, $U(4)$ gauge symmetry arises from the $a$-stack of D6-branes, while the $U(2)_L$ gauge symmetry and the $U(2)_R$ gauge symmetry arise from the 
$b$-stack of D6-branes and $c$-stack of D6-branes, respectively.
The strong, weak and hypercharge gauge couplings $g^2_a,  g^2_b$ and $\frac{5}{3}g^2_Y$ are constricted by the gauge coupling relation~\eqref{gaugecoupling}. 
As it has been mentioned in the introduction, to approach string-scale gauge coupling relation, RGEs evolution methods are often utilized~\cite{Chen:2017rpn,Chen:2018ucf,He:2021kbj}.
In practice, at the two-loop level, the RGEs for the gauge couplings are presented in~\cite{Barger:2004sf,Barger:2007qb, Barger:2005qy,Gogoladze:2010in}. Such that 
{\small \begin{equation}\label{eq:rge}
	\frac{d}{d\ln \mu} g_i=\frac{b_i}{(4\pi)^2}g_i^3 +\frac{g_i^3}{(4\pi)^4}
	\left[ \sum_{j=1}^3 B_{ij}g_j^2-\sum_{\alpha=u,d,e} d_i^\alpha
	{\rm Tr}\left( h^{\alpha \dagger}h^{\alpha}\right) \right],
\end{equation}}
in which $g_i(i=1,2,3)$ represents the SM gauge couplings, while the Yukawa couplings are represented by $h^{\alpha}(\alpha=u,d,e)$. 
And in SM the coefficients for the above beta functions are given in \cite{Machacek:1983tz,Machacek:1983fi,Machacek:1984zw,Cvetic:1998uw} as
{ \begin{eqnarray}\label{eq:rge-sm}
		&&b_{\rm SM}=\left(\frac{41}{6} \frac{1}{k_Y},-\frac{19}{6}\frac{1}{k_2},-7\right) ,~
		B_{\rm SM}=\begin{pmatrix}
			\frac{199}{18} \frac{1}{k_Y^2} &
			\frac{27}{6} \frac{1}{k_Y k_2} &\frac{44}{3} \frac{1}{k_Y} \cr 
			\frac{3}{2} \frac{1}{k_Y k_2} & \frac{35}{6}\frac{1}{k_2^2}&12\frac{1}{k_2} \cr
			\frac{11}{6} \frac{1}{k_Y} &\frac{9}{2}\frac{1}{ k_2}&-26 
		\end{pmatrix},~\\
		&&d^u_{\rm SM}=\left(\frac{17}{6} \frac{1}{k_Y} ,\frac{3}{2}\frac{1}{k_2},2\right),~
		d^d_{\rm SM}=0,~
		d^e_{\rm SM}=0,
\end{eqnarray}}
while in the supersymmetric models the coefficients are given in \cite{Barger:1992ac,Barger:1993gh,Martin:1993zk} by
{ \begin{eqnarray}\label{eq:rge-susy}
		&&b_{\rm SUSY}=\left(11 \frac{1}{k_Y},\frac{1}{k_2},-3\right) ,~ 
		B_{\rm SUSY}=
		\begin{pmatrix}
			\frac{199}{9}
			\frac{1}{k_Y^2}&  9\frac{1}{k_Yk_2}&\frac{88}{3} \frac{1}{k_Y} \cr
			3\frac{1}{k_Yk_2} & 25\frac{1}{k_2^2}&24\frac{1}{k_2} \cr
			\frac{11}{3}\frac{1}{k_Y} & 9\frac{1}{k_2} & 14
		\end{pmatrix},~ \\
		&&d^u_{\rm SUSY}=\left(\frac{26}{3} \frac{1}{k_Y},6\frac{1}{k_2},4\right) ,~
		d^d_{\rm SUSY}=0,~
		d^e_{\rm SUSY}=0,
\end{eqnarray}}
in which $k_Y$ and $k_2$ represent the general normalization factors.

\subsection{String-scale Gauge Coupling Relations}

In order to obtain the two-loop RGEs for SM gauge couplings, one-loop RGEs for Yukawa couplings need to be numerically studied with the new physics contributions and threshold considered~\cite{Gogoladze:2010in}.

To explore the renormalization group running of gauge couplings $g_i$ from the Z-boson mass scale $M_Z$ up to high energies, we numerically solve the two-loop RGEs for SM gauge couplings and one-loop RGEs for Yukawa couplings. The entire evolution is structured into three distinct stages, with each stage characterized by specific particle content and corresponding RGEs. In which, the first stage describes the coupling evolution from $M_Z$ to the supersymmetry breaking scale $M_{\rm S}$.  Within this range, the particle spectrum is dominated by SM fields, such that the running of $g_i(\mu)$ is governed by the non-supersymmetric two-loop RGEs \eqref{eq:rge-sm}. The initial conditions at $\mu=M_Z$ are set by experimental measurements of SM gauge couplings, 
\begin{equation}
	g_1(M_Z)=\sqrt{k_Y}\frac{g_{em}}{\cos\theta_W}~,~g_2(M_Z)=\sqrt{k_2}\frac{g_{em}}{\sin\theta_W}~,~g_3(M_Z)=\sqrt{4\pi \alpha_s}~.
\end{equation}
Solving these RGEs yields the values of $g_i$ at $M_{\rm S}$, which serve as the boundary conditions for the subsequent supersymmetric evolution stage. Moving to the second stage from $M_S$ to $M_{V}$, where $M_V$ denotes the mass threshold of vector-like particles, the particle content expands to include SM superpartners. 
The running of gauge coupling is now dictated by the supersymmetric two-loop RGEs \eqref{eq:rge-susy}, which incorporate contributions from both SM fields and their superpartners. The third stage encompasses the energy interval $M_{V}\leq \mu \leq M_{\rm U}$, where we extend the solution of supersymmetric RGEs to incorporate both one-loop and two-loop quantum corrections from the vector-like particles. Notably, the corrections from these vector-like particles modify the slope of $g_i(\mu)$  running, thereby driving the three gauge couplings toward mutual unification value to the scale $M_{\rm {U}}$. By propagating $g_i{(\mu)}$ through the above three stages, we find that they converge to a unified value $g_{\rm U}$ at $M_{\rm U}$.

The $SU(3)_C\times SU(2)_L\times U(1)_Y$ quantum numbers assigned to the vector-like particles and their contributions to the one-loop beta functions~\cite{Jiang:2006hf,Barger:2007qb} are listed below.
\begin{eqnarray}
	&& XQ + {\overline{XQ}} = {\mathbf{(3, 2, {1\over 6}) + ({\bar 3}, 2,
			-{1\over 6})}}\,, \quad \Delta b =({1\over 5}, 3, 2)\,;\\ 
	&& XU + {\overline{XU}} = {\mathbf{ ({3},
			1, {2\over 3}) + ({\bar 3},  1, -{2\over 3})}}\,, \quad \Delta b =
	({8\over 5}, 0, 1)\,;\\ 
	&& XD + {\overline{XD}} = {\mathbf{ ({3},
			1, -{1\over 3}) + ({\bar 3},  1, {1\over 3})}}\,, \quad \Delta b =
	({2\over 5}, 0, 1)\,;\\  
	&& XL + {\overline{XL}} = {\mathbf{(1,  2, {1\over 2}) + ({1},  2,
			-{1\over 2})}}\,, \quad \Delta b = ({3\over 5}, 1, 0)\,;\\ 
	&& XE + {\overline{XE}} = {\mathbf{({1},  1, {1}) + ({1},  1,
			-{1})}}\,, \quad \Delta b = ({6\over 5}, 0, 0)\,;\\ 
	&& XG = {\mathbf{({8}, 1, 0)}}\,, \quad \Delta b = (0, 0, 3)\,;\\ 
	&& XW = {\mathbf{({1}, 3, 0)}}\,, \quad \Delta b = (0, 2, 0)\,;\\
	&& XT + {\overline{XT}} = {\mathbf{(1, 3, 1) + (1, 3,
			-1)}}\,, \quad \Delta b =({{18}\over 5}, 4, 0)\,;\\ 
	&& XS + {\overline{XS}} = {\mathbf{(6,  1, -{2\over 3}) + ({\bar 6},
			1, {2\over 3})}}\,, \quad \Delta b = ({16\over 5}, 0, 5)\,;\\ 
	&& XY + {\overline{XY}} = {\mathbf{(3, 2, -{5\over 6}) + ({\bar 3}, 2,
			{5\over 6})}}\,, \quad \Delta b =(5, 3, 2)\,. \label{eq:XW} 
\end{eqnarray}
Following the formalism of~\cite{Barger:2007qb}, which provides a complete treatment of two-loop vector-like particle effects in SUSY models, we augment the supersymmetric RGEs \eqref{eq:rge-susy} with these corrections. The full beta functions \eqref{eq:rge} governing the third stage of evolution thus incorporate both one- and two-loop contributions from the vector-like particles:
\begin{equation}
   \beta_i=\beta_i^{(0)}+\beta_i^{(1,\rm VLP)}+\beta_i^{(2,\rm VLP)}~, 
\end{equation}
where $\beta_i^{(0)}$ denotes the supersymmetric beta function following stage 2, and $\beta_i^{(1,\rm VLP)}$, $\beta_i^{(2,\rm VLP)}$ are the one- and two-loop vector-like particle induced corrections, respectively.

To be concrete, consider a vector-like quark pair $XQ+\overline{XQ}$, transforming as $(3,2,1/6)$ under $SU(3)_C\times SU(2)_L\times U(1)_Y$. This pair contributes non-trivially to all three $\Delta b_i$ coefficients, thereby depressing the running of each corresponding gauge coupling $g_i$. The degree of depression increases as $\Delta b_i$ increases. That is to say, the inverse hypercharge coupling $\alpha^{-1}$ decreases more rapidly in models with the up-type vector-like quarks $XU+\overline{XU}$ than that in models with down-type vector-like quarks $XD+\overline{XD}$.
Conversely, the $SU(3)_C$ adjoint $XG$ is a singlet under $SU(2)_L\times U(1)_Y$, resulting in vanishing contributions $\Delta b_{1,2}=0$. Consequently, while the running of $g_1$ and $g_2$ remains unchanged from Stage 2, the strong coupling $g_3$ is selectively adjusted by $XG$. This demonstrates how different vector-like particles can target specific gauge couplings, enabling precise tuning of the unification behavior.

\subsection{Gauge Coupling Relation under Symmetry Breaking}
Gauge coupling evolution requires well-defined initial conditions, vector-like particle content, and RGE structures. In this section, we specify these key points, consistent with experimental constraints and theoretical requirements.

Initial SM parameters at $M_Z$ are fixed using experimental measurements \cite{ParticleDataGroup:2018ovx,ParticleDataGroup:2020ssz,ATLAS:2014wva,dEnterria:2022hzv}, serving as boundary condition for stage 1 evolution from $M_Z$ to $M_S$:
\begin{equation}
	\begin{split}
		&M_Z=91.1876 {\rm~GeV},~
		m_t=173.34\pm 0.27({\rm{stat}})\pm 0.71 ({\rm{syst}}){\rm{~GeV}},~
		v=174.10 {\rm{~GeV}},\\
		&\alpha_s(M_Z)=0.1179\pm 0.0009,~
		\alpha_{em}^{-1}(M_Z)=128.91\pm 0.02,~
		\sin^2\theta_W(M_Z)=0.23122.
	\end{split}
\end{equation}
where  $M_Z$ is Z boson mass, $m_t$ is top quark pole mass,  $v$ is Higgs vacuum expectation value, $\alpha_s$ is strong coupling constant, $\alpha_{em}$ is fine structure constant, and $\theta_W$ is weak mixing angle. The supersymmetric breaking scale $M_{\rm S}\geq 1$ TeV is constrained by lower supersymmetric particle search bounds and gauge hierarchy preservation. Benchmark values $2.5$ TeV or $3.0$ TeV are tested: variations induce $<5\%$ shifts in the unification scale $M_{\rm U}$, with larger $M_{\rm S}$ lowering $M_{\rm U}$. For consistency, $M_{\rm S}=3$ TeV is fixed.

\begin{figure}[!h]\centering
	\includegraphics[width=0.48\linewidth]{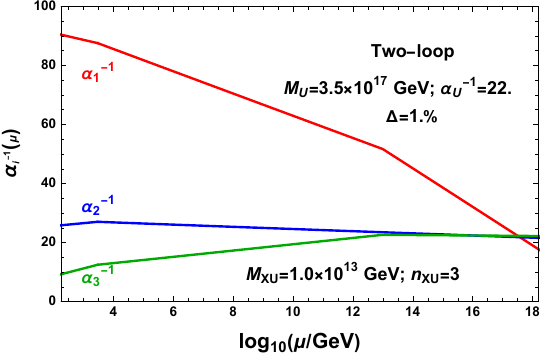}\quad\includegraphics[width=0.48\linewidth]{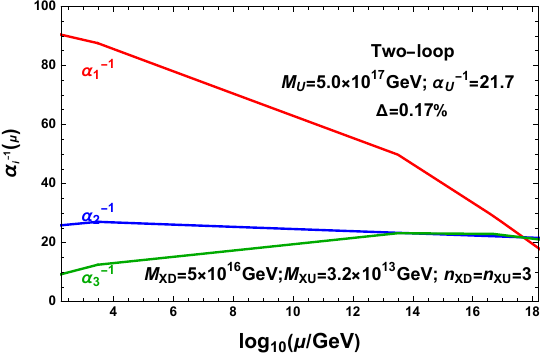}
	\caption{The evolution of two-loop gauge couplings in the Model 1 with vector-like particles $(XU, \overline{XU})$ and $(XD, \overline{XD})$. The masses of these particles are set as $M_{XU}=1.0\times 10^{13}$ GeV (left panel) and $M_{XD}=5\times 10^{16}$ GeV, $M_{XU}=3.2\times 10^{13}$ GeV (right panel).  }\label{fig:model1}
\end{figure}

\begin{figure}[!h]\centering
	\includegraphics[width=0.48\linewidth]{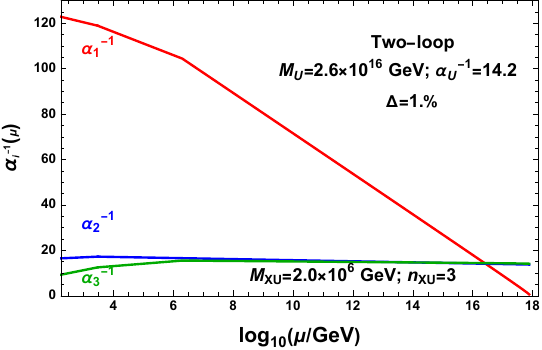}\quad
    \includegraphics[width=0.48\linewidth]{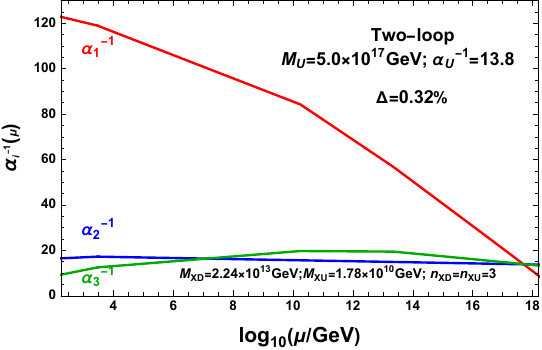}
	\caption{The evolution of two-loop gauge couplings in the Model 2 with vector-like particles $(XU, \overline{XU})$ and $(XD, \overline{XD})$. The masses of these particles are set as $M_{XU}=2.0\times 10^{6}$ GeV (left panel) and $M_{XD}=2.24\times 10^{13}$ GeV, $M_{XU}=1.78\times 10^{10}$ GeV (right panel). }\label{fig:model2}
\end{figure}

\begin{figure}[!h]\centering
	\includegraphics[width=0.48\linewidth]{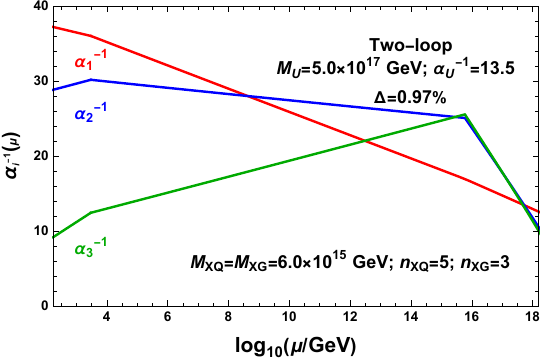}
	\caption{The evolution of two-loop gauge couplings in the Model 3-m with vector-like particles $5(XQ, \overline{XQ})+3XG$. The masses of these particles are set as $M_{XQ}=M_{XG}=6.0\times 10^{15}$ GeV. }\label{fig:model3-m}
\end{figure}

Take Model 1, 2 and Model 3-m\,(with $SU(2)_L$ gauge coupling modified/suppressed by symmetry breaking) as examples, we study their gauge coupling relations at string scale as presented in Figure \ref{fig:model1},\,\ref{fig:model2},\,\ref{fig:model3-m}, in which the couplings are redefined as
\begin{equation}
    \alpha_1\equiv k_Y \frac{g_Y^2}{4\pi}~,~~\alpha_2\equiv k_2 \frac{g_b^2}{4\pi}~,~~\alpha_3\equiv k_3\frac{g_a^2}{4\pi}~.
\end{equation}
where $g_Y,g_b,g_a$ are bare couplings of $U(1)_Y$, $SU(2)_L$ $SU(3)_C$ and $k_Y, k_2,k_3$ are normalization factor. 
Take $k_3=1$, unification at $M_U$ is defined as $\alpha_U^{-1}\equiv \alpha_1^{-1}=\left(\alpha_2^{-1}+\alpha_3^{-1}\right)/2$, where $\alpha_U$ denotes the unified gauge coupling. The relative error $\Delta=|\alpha_1^{-1}-\alpha_2^{-1}|/\alpha_1^{-1}$ is limited to be less than $1.0\%$. 
For the evolution curves of $\alpha_i^{-1}(\mu)$ \textit{vs} $\log_{10}(\mu)$, two distinct kinks emerge as hall mark features: the first arises at $M_{\rm S}$ due to the transition from non-supersymmetric to supersymmetric RGEs, while the second is induced at $M_V$ due to the introduction of vector-like particle contributions. These kinks are experimentally testable in the future high-energy colliders.

For the first two models, we find that with normalized factor $k_Y<1$ and $k_2>1$,  the intersection of $\alpha_1^{-1}$ and $\alpha_3^{-1}$ curves lies above $\alpha_2^{-1}$ curve, preventing the unification of all three gauge couplings. 
To resolve this discrepancy and achieve the simultaneous intersection of $\alpha_i^{-1}$ at $M_{\text{string}}$, we introduce vector-like up-type/down-type quarks, $(XD+\overline{XD})$ or $(XU+\overline{XU})$, from the ${\cal N}=2$ subsector. These particles transform as $(3,1,1/3)$ or $(3,1,2/3)$ under $SU(3)_C\times SU(2)_L\times U(1)_Y$, respectively, and their inclusion adjusts the beta functions governing coupling running. Specifically, their non-trivial quantum numbers contribute to $\Delta b_1$ and $\Delta b_3$, while yielding $\Delta b_2=0$.  This selective modification adjusts the slopes of the $U(1)$ and strong couplings relative to electroweak coupling, enabling the alignment of all three curves at $M_{\text{string}}$. Notably, the degree of slope adjustment depends on the number of vector-like particle pairs $n_V$ and their masses $M_{V}$. Increasing $n_V$ enhances the $\Delta b_1$ and $\Delta b_3$ corrections, while heavier $M_V$ delays their impact until higher energy scales. By tuning these parameters, we numerically demonstrate the convergence of $\alpha_i^{-1}$ at $M_{\text{string}}$.

From brane construction, the number of vector-like particles $(XD+\overline{XD})$ and $(XU+\overline{XU})$ is determined by the intersection number $I_{ac}$ of $a$- and $c$-brane stacks. For Models 1 and 2 (Tables~\ref{model1} and \ref{model2}), $I_{ac}=3$, corresponding to three pairs of vector-like particles. As shown in Figure~\ref{fig:model1}, after fine-tuning their masses, unification in Model 1 is achieved at $M_U=3.5\times 10^{17}$ GeV by introducing $(XU,\overline{XU})$ at $M_{XU}=1.0\times 10^{13}$ GeV. If both $(XD+\overline{XD})$ and $(XU+\overline{XU})$ are included, the precise string scale relation is achieved at $M_U=5.0\times 10^{17}$ GeV, with $M_{XD}=5\times 10^{16}$ GeV and $M_{XU}=3.2\times 10^{13}$ GeV. Moreover, three pairs of $(XD,~\overline{XD})$ and $(XU, ~\overline{XU})$  can couple to the nine Higgs fields in the symmetric representation of $SU(2)_R$. The $SU(2)_R$ gauge symmetry is broken down to the $U(1)_{I_{3R}}$ gauge symmetry via D6-brane splitting, and then the Yukawa couplings for  $(XD,~\overline{XD})$ and $(XU,~\overline{XU})$ can be  different. 
While the Yukawa couplings for $(XD,~\overline{XD})$ are about three orders larger than those for
$(XU,~\overline{XU})$, we obtain a significant mass splitting between $XD$ and $XU$ which is crucial to achieve precise gauge coupling relation, as it allows independent tuning of $U(1)$ and strong coupling across different energy scales.

For Model 2, the factors $k_2=11/6$ and $k_Y=5/14$ deviate significantly from 1, resulting in an  initial value of electroweak coupling that lies very close to strong coupling at $M_Z$. Consequently, even with the inclusion of 3 $(XU,\overline{XU})$ pairs at $M_{XU}=2.0\times 10^{6}$ GeV, the unification scale remains slightly below the string scale, at $M_{\text{U}}=2.6\times 10^{16}$ GeV. 
To further push the unification energy scale toward $M_{\text{string}}$, we introduce additional $(XD,\overline{XD})$ pairs, which are integrated into the RGE evolution at two distinct thresholds $M_{XD}=2.24\times 10^{13}$ GeV and $M_{XU}=1.78\times 10^{10}$ GeV. As shown in the right panel of Figure~\ref{fig:model2}, three kinks are visible in the $\alpha_1^{-1}$ and $\alpha_3^{-1}$ curves, corresponding to the successive integration of $(XD+\overline{XD})$ and $(XU,\overline{XU})$. This multi-step inclusion enhances the $\Delta_1$ and $\Delta_3$ corrections above the corresponding thresholds, gradually adjusting the slops of $\alpha_1^{-1}$ and $\alpha_3^{-1}$. Through this hierarchical integration, we achieve  precise gauge coupling at $5.0\times 10^{17}$ GeV.

In the case of Model 3, the large constant $k_2$ in the gauge coupling relation makes it difficult to achieve unification at the target energy scale through renormalization group running.
Since $k_2>1$, this implies that at the unification scale the value of $\alpha_2^{-1}$ must be smaller than $\alpha_3^{-1}$.
Under renormalization group running, the curve of $\alpha_2^{-1}$ typically lies between those of $\alpha_1^{-1}$ and $\alpha_3^{-1}$.
It is very difficult to make it ``drop'' below the $\alpha_3^{-1}$ curve, let alone fall significantly lower.

Fortunately, stems from the canonical normalization requirement of the gauge kinetic terms get modified after symmetry breaking as shown in Section~\ref{sec:symm}.
The $\grSU(2)_{L_1}\times\grSU(2)_{L_2}$ symmetry can be broken down to the diagonal subgroup $\grSU(2)_{L'}$ through the mechanism of vacuum expectation values.
The coupling constant $g_{L'}$ of the resulting diagonal subgroup $\grSU(2)$ after symmetry breaking get suppressed which provides an alternative approach to achieve string-scale gauge coupling unification.

As shown in Figure~\ref{fig:model3-m},
with $k_2$ lowered via this approach, gauge coupling unification at the desired energy level becomes easier, with vector-like particles $(XQ+\overline{XQ})$ and $XG$ with masses $M_{XQ}=M_{XG}=6.0\times 10^{15}$ introduced. The vector-like particles $(XQ,\overline{XQ})$ arise from the intersections between $a$- and $b$-stacks of D6-brane. Three chiral multiplets $XG$ arise from \textit{aa} sector of the adjoint representation of $SU(4)_C$.
The number of introduced particle $(XQ,\overline{XQ})$ is determined by the intersection number of $a$- and $b$-brane stacks $I_{ab}=5$, and therefore $n_{XQ}=5$.
From this approach, the slope of $\alpha_2^{-1}$ and $\alpha_3^{-1}$ is further suppressed than $\alpha_1^{-1}$, and thus we achieve unification at string-scale, $M_{U}=5.0\times 10^{17}$ GeV.

Based on the above investigation, here we find the string-scale gauge coupling relation can be achieved at two-loop level via introducing the extra vector-like particles from ${\cal N}=2$ subsector in these models. 
Furthermore, for the constructed models with large $k_2$, by utilizing the symmetry breaking mechanism introduced in  Section~\ref{sec:symm}, we propose that the large  $k_2$ can be highly suppressed and thus string-scale gauge coupling unification can be realized in a good manner as well.
Again, the quantum numbers for these vector-like particles are the same as those of the SM fermions and their Hermitian conjugates. The number $n_V$ of these particles is highly model dependent and can be determined from the corresponding brane intersection number. 
As discussed in Ref.~\cite{Li:2022cqk}, these additional particles can be decoupled through Higgs mechanism or instanton effects. 
With multiple pairs of extra particles emerge naturally, one only needs to fine-tune their masses to unify the gauge coupling near the string scale 
\begin{equation}
    M_{\text U}\sim M_{\text{string}}\simeq 5\times 10^{17}\text{~GeV}.
\end{equation} 
To summarize, in our numerical check with symmetry breaking considered, we observe that the masses of these extra particles 
are determined by three factors: (i) the number of the particles, (ii) the beta functions and (iii) the precise energy scale where we realize the string-scale gauge coupling relations. Moreover, as $n_V$ increases, the masses of these extra particles
approach $M_{\text{string}}$.

\section{Conclusions and Outlook}

In this paper, realizing that the three-family of chiral fermion condition can actually be generalized from 
$I_{ab}+I_{ab'}=3~; \,I_{ac}=-3, \,I_{ac'}=0$\,(or $I_{ac}=0,\, I_{ac'}=-3$)
to $I_{ab}+I_{ab'}=3,\, I_{ac}+I_{ac'}=-3$,
we in particular searched for the new supersymmetric Pati-Salam models with the systematic algorithm built in~\cite{He:2021gug}.
While three generations of particles are realized by the brane intersection of $a$- and $b/b'$- stacks of branes, with $I_{ac}+I_{ac'}=-3$, four classes of new supersymmetric Pati-Salam models were obtained, represented by Model 1,\,2,\,3,\,4.
As follows, whether these new models arisen from the extended three-family of chiral fermions condition, such as $I_{ac}=3, \,I_{ac'}=-6$ and $I_{ac}=-1,\, I_{ac'}=-2$, still preserve the string-scale gauge coupling unification are also investigated with symmetry breaking mechanism imposed. 

Firstly, note that the three generation condition in supersymmetric Pati-Salam models can be extended and the three-family of chiral fermions can arise from different brane intersection, which further extend the landscape of ~$\mathcal{N}=1$ supersymmetric Pati-Salam models that found in~\cite{He:2021gug}.
From the former investigation, these models are speculated to be very limited~\cite{Cvetic:2004ui}. Nevertheless, in this paper we generalized the supersymmetric Pati-Salam models search to this scope. 
We in particular focused on the extended model building of supersymmetric Pati-Salam models from brane intersection. 
With the complete search method, we found there are four new classes of supersymmetric Pati-Salam models that were omitted under the former three generation condition. In precise, the three generation condition can provide three-family of chiral fermions with $I_{ac}=3, I_{ac'}=-6$ and $I_{ac}=-1, I_{ac'}=-2$ as well. And for each above three generation condition, there are two classes of physical independent models, each with 6144 number of new supersymmetric Pati-Salam models found\footnote{Examples of these new supersymmetric Pati-Salam models with duality variations are presented in the Appendix.}. 
These new models are in principle found via iteratively solving the Diophantine equations constituting of the new three generation conditions, RR tadpole cancellation conditions, and supersymmetry conditions.
In each class, the supersymmetric Pati-Salam models have the same gauge coupling relation, while with distinct gauge coupling relations from the former investigation~\cite{He:2021gug}.
Take representative models from each class as example, we phenomenologically studied the spectrum for each class. The Higgs multiples arise differently from the former investigations. In particular, these extended part of the supersymmetric Pati-Salam models are with rare filler branes, and maximally with two filler branes. 

Secondly, consider the symmetry breaking method to suppress the $SU(2)_L$ gauge coupling, we show that the symmetry breaking mechanism helps to achieve string-scale gauge coupling relation at around  $M_{\text{string}}$. This symmetry breaking mechanism is especially effective for models with high $k_2=g_b^2/g_a^2$, that is often found to be difficult to realize gauge unification in RGEs evolution. 
Our key strategy is to introduce exotic vector-like particles to modify beta functions. We implement this strategy differently across models, matching exotic particles to each model’s specific structure. For Model 1 and 2, we introduce vector-like down-type quarks $(XD + \overline{XD})$ and up-type quarks $(XU + \overline{XU})$, both originating from the $\mathcal{N}=2$ subsector. These particles  give $\Delta b_1$ and $\Delta b_3$ corrections, adjusting the slopes of $\alpha_1^{-1}$ and $\alpha_3^{-1}$ to align with $\alpha_2^{-1}$ at $M_{\text{string}}$. For Model 3-m, with symmetry breaking mechanism imposed, we introduce two complementary particles: vector-like quarks $(XQ + \overline{XQ})$ and a vector-like $SU(4)_C$-adjoint particle $XG$. In particular, the vector-like particles $(XQ,\overline{XQ})$ arise from the intersections between $a$- and $b$-stacks of D6-brane. Three chiral multiplets $XG$ arise from \textit{aa} sector of the adjoint representation of $SU(4)_C$.
Through two-loop RGE analysis, we demonstrate that these particles modify the beta functions to align all gauge couplings at $M_{\text{string}}$ in supersymmetric Pati-Salam models studied. 

Therefore, we proposed to extend the landscape of ${\cal N}=1$ supersymmetric Pati-Salam models with the generalized three-family condition. And with symmetry breaking mechanism imposed, the revised RGE evolution was shown to realize the string-scale gauge coupling relation in an effective manner. Concrete models with string-scale gauge coupling relations achieved for new supersymmetric Pati-Salam models were presented.

For future investigation, we think it will be interesting and worthwhile to study the general three-family of chiral fermion condition, especially in brane intersection models. Although it was shown to be very difficult to realize these models as they are highly constraint by brane intersection, 
with the complete search method constructed in~\cite{He:2021gug} these models can be effectively found. And imposing symmetry breaking mechanism before the RGE evolution provides an effective approach to realize string-scale gauge coupling relation in brane intersection theories.
This provides a robust mechanism for resolving the scale mismatch between GUTs and string theory in generic intersecting D-brane constructions.

\begin{acknowledgments}
    We acknowledge Haotian Huangfu for useful discussion.
	TL is supported in part by the National Key Research and Development Program of China Grant No. 2020YFC2201504, by the Projects No. 11875062, No. 11947302, No. 12047503, and No. 12275333 supported by the National Natural Science Foundation of China, by the Key Research Program of the Chinese Academy of Sciences, Grant No. XDPB15, by the Scientific Instrument Developing Project of the Chinese Academy of Sciences, Grant No. YJKYYQ20190049, by the International Partnership Program of Chinese Academy of Sciences for Grand Challenges, Grant No. 112311KYSB20210012, and by the Henan Province Outstanding Foreign Scientist Studio Project, No.GZS2025008.	 
    RS is supported by the Fundamental Research Funds for the Central Universities Grant No.~E4EQ0102X2.   LW is supported by the Natural Science Basic Research Program of Shaanxi, Grant No. 2024JC-YBMS-039.
	
\end{acknowledgments}



\bibliography{reference}

\appendix

\section{Generalized Supersymmetric Pati-Salam Models}
\label{appA}
\FloatBarrier

In this section, we present some other examples of new supersymmetric Pati-Salam models from the three-family generalization,
which are related to  Model 1\,(Table~\ref{tb:model5}-\ref{tb:model14}), Model 2\,(Table~\ref{tb:model15}-\ref{tb:model24}), 
Model 3\,(Table~\ref{tb:model25}-\ref{tb:model34}), 
Model 4\,(Table~\ref{tb:model35}-\ref{tb:model44}), respectively.
These new supersymmetric Pati-Salam models with the same MSSM gauge coupling relation are related by symmetry transformations, such as, D6-brane Sign Equivalent Principle\,(DSEP), Type I and II T-dualities, as discussed  in~\cite{Cvetic:2004ui, He:2021gug}.

\begin{table}[!h]\scriptsize
	\caption{D6-brane configurations and intersection numbers of Model 5, and its MSSM gauge coupling relation is $g^2_a=\frac{7}{6}g^2_b=\frac{35}{66}g^2_c=\frac{175}{268}(\frac{5}{3}g^2_Y)=\frac{8 \sqrt[4]{2} 5^{3/4} \pi  e^{\phi_4}}{11 \sqrt{3}}$.}
	\label{tb:model5}
	\begin{center}
		\begin{tabular}{|c||c|c||c|c|c|c|c|c|c|}
			\hline\rm{Model} 5 & \multicolumn{9}{c|}{$U(4)\times U(2)_L\times U(2)_R\times USp(2) $}\\
			\hline \hline			\rm{stack} & $N$ & $(n^1,l^1)\times(n^2,l^2)\times(n^3,l^3)$ & $n_{\Ysymm}$& $n_{\Yasymm}$ & $b$ & $b'$ & $c$ & $c'$ & 1\\
			\hline
			$a$ & 8 & $(1,1)\times (1,0)\times (1,-1)$ & 0 & 0  & 3 & 0 & -6 & 3 & 0\\
			$b$ & 4 & $(1,-1)\times (1,-1)\times (1,2)$ & -2 & -6  & - & - & 0 & 8 & 2\\
			$c$ & 4 & $(1,5)\times (0,1)\times (-1,-2)$ & -9 & 9  & - & - & - & - & 10\\
			\hline
			1 & 2 & $(1, 0)\times (1, 0)\times (2, 0)$& \multicolumn{7}{c|}{$x_A = \frac{1}{12}x_B = \frac{1}{10}x_C = \frac{1}{12}x_D$}\\
			& & & \multicolumn{7}{c|}{$\beta^g_1=6$}\\
			& & & \multicolumn{7}{c|}{$\chi_1=\frac{1}{\sqrt{10}}$, $\chi_2=\frac{\sqrt{\frac{5}{2}}}{6}$, $\chi_3=\sqrt{\frac{2}{5}}$}\\
			\hline
		\end{tabular}
	\end{center}
\end{table}

\begin{table}[!h]\scriptsize
	\caption{D6-brane configurations and intersection numbers of Model 6, and its MSSM gauge coupling relation is $g^2_a=\frac{7}{6}g^2_b=\frac{35}{66}g^2_c=\frac{175}{268}(\frac{5}{3}g^2_Y)=\frac{8 \sqrt[4]{2} 5^{3/4} \pi  e^{\phi_4}}{11 \sqrt{3}}$.}
	\label{tb:model6}
	\begin{center}
		\begin{tabular}{|c||c|c||c|c|c|c|c|c|c|}
			\hline\rm{Model} 6 & \multicolumn{9}{c|}{$U(4)\times U(2)_L\times U(2)_R\times USp(2) $}\\
			\hline \hline			\rm{stack} & $N$ & $(n^1,l^1)\times(n^2,l^2)\times(n^3,l^3)$ & $n_{\Ysymm}$& $n_{\Yasymm}$ & $b$ & $b'$ & $c$ & $c'$ & 1\\
			\hline
			$a$ & 8 & $(1,1)\times (1,0)\times (1,-1)$ & 0 & 0  & 3 & 0 & -6 & 3 & 0\\
			$b$ & 4 & $(1,-1)\times (1,-1)\times (1,2)$ & -2 & -6  & - & - & 0 & 8 & 2\\
			$c$ & 4 & $(-1,-5)\times (0,1)\times (1,2)$ & -9 & 9  & - & - & - & - & 10\\
			\hline
			1 & 2 & $(1, 0)\times (1, 0)\times (2, 0)$& \multicolumn{7}{c|}{$x_A = \frac{1}{12}x_B = \frac{1}{10}x_C = \frac{1}{12}x_D$}\\
			& & & \multicolumn{7}{c|}{$\beta^g_1=6$}\\
			& & & \multicolumn{7}{c|}{$\chi_1=\frac{1}{\sqrt{10}}$, $\chi_2=\frac{\sqrt{\frac{5}{2}}}{6}$, $\chi_3=\sqrt{\frac{2}{5}}$}\\
			\hline
		\end{tabular}
	\end{center}
\end{table}

\begin{table}\scriptsize
	\caption{D6-brane configurations and intersection numbers of Model 7, and its MSSM gauge coupling relation is $g^2_a=\frac{7}{6}g^2_b=\frac{35}{66}g^2_c=\frac{175}{268}(\frac{5}{3}g^2_Y)=\frac{8 \sqrt[4]{2} 5^{3/4} \pi  e^{\phi_4}}{11 \sqrt{3}}$.}
	\label{tb:model7}
	\begin{center}
		\begin{tabular}{|c||c|c||c|c|c|c|c|c|c|}
			\hline\rm{Model} 7 & \multicolumn{9}{c|}{$U(4)\times U(2)_L\times U(2)_R\times USp(2) $}\\
			\hline \hline			\rm{stack} & $N$ & $(n^1,l^1)\times(n^2,l^2)\times(n^3,l^3)$ & $n_{\Ysymm}$& $n_{\Yasymm}$ & $b$ & $b'$ & $c$ & $c'$ & 1\\
			\hline
			$a$ & 8 & $(1,1)\times (1,0)\times (1,-1)$ & 0 & 0  & 3 & 0 & 3 & -6 & 0\\
			$b$ & 4 & $(1,-1)\times (1,-1)\times (1,2)$ & -2 & -6  & - & - & 8 & 0 & 2\\
			$c$ & 4 & $(1,-5)\times (0,1)\times (1,-2)$ & 9 & -9  & - & - & - & - & 10\\
			\hline
			1 & 2 & $(1, 0)\times (1, 0)\times (2, 0)$& \multicolumn{7}{c|}{$x_A = \frac{1}{12}x_B = \frac{1}{10}x_C = \frac{1}{12}x_D$}\\
			& & & \multicolumn{7}{c|}{$\beta^g_1=6$}\\
			& & & \multicolumn{7}{c|}{$\chi_1=\frac{1}{\sqrt{10}}$, $\chi_2=\frac{\sqrt{\frac{5}{2}}}{6}$, $\chi_3=\sqrt{\frac{2}{5}}$}\\
			\hline
		\end{tabular}
	\end{center}
\end{table}

\begin{table}\scriptsize
	\caption{D6-brane configurations and intersection numbers of Model 8, and its MSSM gauge coupling relation is $g^2_a=\frac{7}{6}g^2_b=\frac{35}{66}g^2_c=\frac{175}{268}(\frac{5}{3}g^2_Y)=\frac{8 \sqrt[4]{2} 5^{3/4} \pi  e^{\phi_4}}{11 \sqrt{3}}$.}
	\label{tb:model8}  
	\begin{center}
		\begin{tabular}{|c||c|c||c|c|c|c|c|c|c|}
			\hline\rm{Model} 8 & \multicolumn{9}{c|}{$U(4)\times U(2)_L\times U(2)_R\times USp(2) $}\\
			\hline \hline			\rm{stack} & $N$ & $(n^1,l^1)\times(n^2,l^2)\times(n^3,l^3)$ & $n_{\Ysymm}$& $n_{\Yasymm}$ & $b$ & $b'$ & $c$ & $c'$ & 1\\
			\hline
			$a$ & 8 & $(1,1)\times (1,0)\times (1,-1)$ & 0 & 0  & 3 & 0 & -6 & 3 & 0\\
			$b$ & 4 & $(1,-1)\times (1,-1)\times (1,2)$ & -2 & -6  & - & - & 0 & 8 & 2\\
			$c$ & 4 & $(1,5)\times (0,-1)\times (1,2)$ & -9 & 9  & - & - & - & - & 10\\
			\hline
			1 & 2 & $(1, 0)\times (1, 0)\times (2, 0)$& \multicolumn{7}{c|}{$x_A = \frac{1}{12}x_B = \frac{1}{10}x_C = \frac{1}{12}x_D$}\\
			& & & \multicolumn{7}{c|}{$\beta^g_1=6$}\\
			& & & \multicolumn{7}{c|}{$\chi_1=\frac{1}{\sqrt{10}}$, $\chi_2=\frac{\sqrt{\frac{5}{2}}}{6}$, $\chi_3=\sqrt{\frac{2}{5}}$}\\
			\hline
		\end{tabular}
	\end{center}
\end{table}

\begin{table}\scriptsize
	\caption{D6-brane configurations and intersection numbers of Model 9, and its MSSM gauge coupling relation is $g^2_a=\frac{7}{6}g^2_b=\frac{35}{66}g^2_c=\frac{175}{268}(\frac{5}{3}g^2_Y)=\frac{8 \sqrt[4]{2} 5^{3/4} \pi  e^{\phi_4}}{11 \sqrt{3}}$.}
	\label{tb:model9}
	\begin{center}
		\begin{tabular}{|c||c|c||c|c|c|c|c|c|c|}
			\hline\rm{Model} 9 & \multicolumn{9}{c|}{$U(4)\times U(2)_L\times U(2)_R\times USp(2) $}\\
			\hline \hline			\rm{stack} & $N$ & $(n^1,l^1)\times(n^2,l^2)\times(n^3,l^3)$ & $n_{\Ysymm}$& $n_{\Yasymm}$ & $b$ & $b'$ & $c$ & $c'$ & 1\\
			\hline
			$a$ & 8 & $(1,1)\times (1,0)\times (1,-1)$ & 0 & 0  & 3 & 0 & 3 & -6 & 0\\
			$b$ & 4 & $(1,-1)\times (1,-1)\times (1,2)$ & -2 & -6  & - & - & 8 & 0 & 2\\
			$c$ & 4 & $(-1,5)\times (0,-1)\times (1,-2)$ & 9 & -9  & - & - & - & - & 10\\
			\hline
			1 & 2 & $(1, 0)\times (1, 0)\times (2, 0)$& \multicolumn{7}{c|}{$x_A = \frac{1}{12}x_B = \frac{1}{10}x_C = \frac{1}{12}x_D$}\\
			& & & \multicolumn{7}{c|}{$\beta^g_1=6$}\\
			& & & \multicolumn{7}{c|}{$\chi_1=\frac{1}{\sqrt{10}}$, $\chi_2=\frac{\sqrt{\frac{5}{2}}}{6}$, $\chi_3=\sqrt{\frac{2}{5}}$}\\
			\hline
		\end{tabular}
	\end{center}
\end{table}

\begin{table}\scriptsize
	\caption{D6-brane configurations and intersection numbers of Model 10, and its MSSM gauge coupling relation is $g^2_a=\frac{7}{6}g^2_b=\frac{35}{66}g^2_c=\frac{175}{268}(\frac{5}{3}g^2_Y)=\frac{8 \sqrt[4]{2} 5^{3/4} \pi  e^{\phi_4}}{11 \sqrt{3}}$.}
	\label{tb:model10}
	\begin{center}
		\begin{tabular}{|c||c|c||c|c|c|c|c|c|c|}
			\hline\rm{Model} 10 & \multicolumn{9}{c|}{$U(4)\times U(2)_L\times U(2)_R\times USp(2) $}\\
			\hline \hline			\rm{stack} & $N$ & $(n^1,l^1)\times(n^2,l^2)\times(n^3,l^3)$ & $n_{\Ysymm}$& $n_{\Yasymm}$ & $b$ & $b'$ & $c$ & $c'$ & 1\\
			\hline
			$a$ & 8 & $(1,1)\times (1,0)\times (1,-1)$ & 0 & 0  & 3 & 0 & 3 & -6 & 0\\
			$b$ & 4 & $(1,-1)\times (1,-1)\times (1,2)$ & -2 & -6  & - & - & 8 & 0 & 2\\
			$c$ & 4 & $(1,-5)\times (0,-1)\times (-1,2)$ & 9 & -9  & - & - & - & - & 10\\
			\hline
			1 & 2 & $(1, 0)\times (1, 0)\times (2, 0)$& \multicolumn{7}{c|}{$x_A = \frac{1}{12}x_B = \frac{1}{10}x_C = \frac{1}{12}x_D$}\\
			& & & \multicolumn{7}{c|}{$\beta^g_1=6$}\\
			& & & \multicolumn{7}{c|}{$\chi_1=\frac{1}{\sqrt{10}}$, $\chi_2=\frac{\sqrt{\frac{5}{2}}}{6}$, $\chi_3=\sqrt{\frac{2}{5}}$}\\
			\hline
		\end{tabular}
	\end{center}
\end{table}

\begin{table}\scriptsize
	\caption{D6-brane configurations and intersection numbers of Model 11, and its MSSM gauge coupling relation is $g^2_a=\frac{7}{6}g^2_b=\frac{35}{66}g^2_c=\frac{175}{268}(\frac{5}{3}g^2_Y)=\frac{8 \sqrt[4]{2} 5^{3/4} \pi  e^{\phi_4}}{11 \sqrt{3}}$.}
	\label{tb:model11}
	\begin{center}
		\begin{tabular}{|c||c|c||c|c|c|c|c|c|c|}
			\hline\rm{Model} 11 & \multicolumn{9}{c|}{$U(4)\times U(2)_L\times U(2)_R\times USp(2) $}\\
			\hline \hline			\rm{stack} & $N$ & $(n^1,l^1)\times(n^2,l^2)\times(n^3,l^3)$ & $n_{\Ysymm}$& $n_{\Yasymm}$ & $b$ & $b'$ & $c$ & $c'$ & 1\\
			\hline
			$a$ & 8 & $(1,1)\times (1,0)\times (1,-1)$ & 0 & 0  & 3 & 0 & -6 & 3 & 0\\
			$b$ & 4 & $(1,-1)\times (1,-1)\times (1,2)$ & -2 & -6  & - & - & 0 & 8 & 2\\
			$c$ & 4 & $(-1,-5)\times (0,-1)\times (-1,-2)$ & -9 & 9  & - & - & - & - & 10\\
			\hline
			1 & 2 & $(1, 0)\times (1, 0)\times (2, 0)$& \multicolumn{7}{c|}{$x_A = \frac{1}{12}x_B = \frac{1}{10}x_C = \frac{1}{12}x_D$}\\
			& & & \multicolumn{7}{c|}{$\beta^g_1=6$}\\
			& & & \multicolumn{7}{c|}{$\chi_1=\frac{1}{\sqrt{10}}$, $\chi_2=\frac{\sqrt{\frac{5}{2}}}{6}$, $\chi_3=\sqrt{\frac{2}{5}}$}\\
			\hline
		\end{tabular}
	\end{center}
\end{table}

\begin{table}\scriptsize
	\caption{D6-brane configurations and intersection numbers of Model 12, and its MSSM gauge coupling relation is $g^2_a=\frac{7}{6}g^2_b=\frac{35}{66}g^2_c=\frac{175}{268}(\frac{5}{3}g^2_Y)=\frac{8 \sqrt[4]{2} 5^{3/4} \pi  e^{\phi_4}}{11 \sqrt{3}}$.}
	\label{tb:model12}
	\begin{center}
		\begin{tabular}{|c||c|c||c|c|c|c|c|c|c|}
			\hline\rm{Model} 12 & \multicolumn{9}{c|}{$U(4)\times U(2)_L\times U(2)_R\times USp(2) $}\\
			\hline \hline			\rm{stack} & $N$ & $(n^1,l^1)\times(n^2,l^2)\times(n^3,l^3)$ & $n_{\Ysymm}$& $n_{\Yasymm}$ & $b$ & $b'$ & $c$ & $c'$ & 1\\
			\hline
			$a$ & 8 & $(1,1)\times (1,0)\times (1,-1)$ & 0 & 0  & 0 & 3 & 3 & -6 & 0\\
			$b$ & 4 & $(1,1)\times (1,1)\times (1,-2)$ & 2 & 6  & - & - & 0 & -8 & 2\\
			$c$ & 4 & $(-1,5)\times (0,1)\times (-1,2)$ & 9 & -9  & - & - & - & - & 10\\
			\hline
			1 & 2 & $(1, 0)\times (1, 0)\times (2, 0)$& \multicolumn{7}{c|}{$x_A = \frac{1}{12}x_B = \frac{1}{10}x_C = \frac{1}{12}x_D$}\\
			& & & \multicolumn{7}{c|}{$\beta^g_1=6$}\\
			& & & \multicolumn{7}{c|}{$\chi_1=\frac{1}{\sqrt{10}}$, $\chi_2=\frac{\sqrt{\frac{5}{2}}}{6}$, $\chi_3=\sqrt{\frac{2}{5}}$}\\
			\hline
		\end{tabular}
	\end{center}
\end{table}

\begin{table}\scriptsize
	\caption{D6-brane configurations and intersection numbers of Model 13, and its MSSM gauge coupling relation is $g^2_a=\frac{7}{6}g^2_b=\frac{35}{66}g^2_c=\frac{175}{268}(\frac{5}{3}g^2_Y)=\frac{8 \sqrt[4]{2} 5^{3/4} \pi  e^{\phi_4}}{11 \sqrt{3}}$.}
	\label{tb:model13}
	\begin{center}
		\begin{tabular}{|c||c|c||c|c|c|c|c|c|c|}
			\hline\rm{Model} 13 & \multicolumn{9}{c|}{$U(4)\times U(2)_L\times U(2)_R\times USp(2) $}\\
			\hline \hline			\rm{stack} & $N$ & $(n^1,l^1)\times(n^2,l^2)\times(n^3,l^3)$ & $n_{\Ysymm}$& $n_{\Yasymm}$ & $b$ & $b'$ & $c$ & $c'$ & 1\\
			\hline
			$a$ & 8 & $(1,1)\times (1,0)\times (1,-1)$ & 0 & 0  & 0 & 3 & -6 & 3 & 0\\
			$b$ & 4 & $(1,1)\times (1,1)\times (1,-2)$ & 2 & 6  & - & - & -8 & 0 & 2\\
			$c$ & 4 & $(1,5)\times (0,1)\times (-1,-2)$ & -9 & 9  & - & - & - & - & 10\\
			\hline
			1 & 2 & $(1, 0)\times (1, 0)\times (2, 0)$& \multicolumn{7}{c|}{$x_A = \frac{1}{12}x_B = \frac{1}{10}x_C = \frac{1}{12}x_D$}\\
			& & & \multicolumn{7}{c|}{$\beta^g_1=6$}\\
			& & & \multicolumn{7}{c|}{$\chi_1=\frac{1}{\sqrt{10}}$, $\chi_2=\frac{\sqrt{\frac{5}{2}}}{6}$, $\chi_3=\sqrt{\frac{2}{5}}$}\\
			\hline
		\end{tabular}
	\end{center}
\end{table}

\begin{table}\scriptsize
	\caption{D6-brane configurations and intersection numbers of Model 14, and its MSSM gauge coupling relation is $g^2_a=\frac{7}{6}g^2_b=\frac{35}{66}g^2_c=\frac{175}{268}(\frac{5}{3}g^2_Y)=\frac{8 \sqrt[4]{2} 5^{3/4} \pi  e^{\phi_4}}{11 \sqrt{3}}$.}
	\label{tb:model14}
	\begin{center}
		\begin{tabular}{|c||c|c||c|c|c|c|c|c|c|}
			\hline\rm{Model} 14 & \multicolumn{9}{c|}{$U(4)\times U(2)_L\times U(2)_R\times USp(2) $}\\
			\hline \hline			\rm{stack} & $N$ & $(n^1,l^1)\times(n^2,l^2)\times(n^3,l^3)$ & $n_{\Ysymm}$& $n_{\Yasymm}$ & $b$ & $b'$ & $c$ & $c'$ & 1\\
			\hline
			$a$ & 8 & $(1,1)\times (1,0)\times (1,-1)$ & 0 & 0  & 0 & 3 & -6 & 3 & 0\\
			$b$ & 4 & $(1,1)\times (1,1)\times (1,-2)$ & 2 & 6  & - & - & -8 & 0 & 2\\
			$c$ & 4 & $(-1,-5)\times (0,1)\times (1,2)$ & -9 & 9  & - & - & - & - & 10\\
			\hline
			1 & 2 & $(1, 0)\times (1, 0)\times (2, 0)$& \multicolumn{7}{c|}{$x_A = \frac{1}{12}x_B = \frac{1}{10}x_C = \frac{1}{12}x_D$}\\
			& & & \multicolumn{7}{c|}{$\beta^g_1=6$}\\
			& & & \multicolumn{7}{c|}{$\chi_1=\frac{1}{\sqrt{10}}$, $\chi_2=\frac{\sqrt{\frac{5}{2}}}{6}$, $\chi_3=\sqrt{\frac{2}{5}}$}\\
			\hline
		\end{tabular}
	\end{center}
\end{table}

\begin{table}\scriptsize
	\caption{D6-brane configurations and intersection numbers of Model 15, and its MSSM gauge coupling relation is $g^2_a=\frac{11}{6}g^2_b=\frac{5}{14}g^2_c=\frac{25}{52}(\frac{5}{3}g^2_Y)=\frac{8}{63} 5^{3/4} \sqrt{11} \pi  e^{\phi_4}$.}
	\label{tb:model15}
	\begin{center}
		\begin{tabular}{|c||c|c||c|c|c|c|c|c|c|c|}
			\hline\rm{Model} 15 & \multicolumn{10}{c|}{$U(4)\times U(2)_L\times U(2)_R\times USp(2)^2 $}\\
			\hline \hline			\rm{stack} & $N$ & $(n^1,l^1)\times(n^2,l^2)\times(n^3,l^3)$ & $n_{\Ysymm}$& $n_{\Yasymm}$ & $b$ & $b'$ & $c$ & $c'$ & 1 & 4\\
			\hline
			$a$ & 8 & $(-1,1)\times (-1,1)\times (-1,1)$ & 0 & 4  & 3 & 0 & 3 & -6 & 1 & 1\\
			$b$ & 4 & $(2,1)\times (0,1)\times (-1,-1)$ & 1 & -1  & - & - & 10 & -9 & 1 & 0\\
			$c$ & 4 & $(-1,2)\times (-1,0)\times (5,1)$ & 9 & -9  & - & - & - & - & 0 & 1\\
			\hline
			1 & 2 & $(1, 0)\times (1, 0)\times (2, 0)$& \multicolumn{8}{c|}{$x_A = 22x_B = 2x_C = \frac{11}{5}x_D$}\\
			4 & 2 & $(0, 1)\times (0, 1)\times (2, 0)$& \multicolumn{8}{c|}{$\beta^g_1=-3$, $\beta^g_4=-3$}\\
			& & & \multicolumn{8}{c|}{$\chi_1=\frac{1}{\sqrt{5}}$, $\chi_2=\frac{11}{\sqrt{5}}$, $\chi_3=4 \sqrt{5}$}\\
			\hline
		\end{tabular}
	\end{center}
\end{table}

\begin{table}\scriptsize
	\caption{D6-brane configurations and intersection numbers of Model 16, and its MSSM gauge coupling relation is $g^2_a=\frac{11}{6}g^2_b=\frac{5}{14}g^2_c=\frac{25}{52}(\frac{5}{3}g^2_Y)=\frac{8}{63} 5^{3/4} \sqrt{11} \pi  e^{\phi_4}$.}
	\label{tb:model16}
	\begin{center}
		\begin{tabular}{|c||c|c||c|c|c|c|c|c|c|c|}
			\hline\rm{Model} 16 & \multicolumn{10}{c|}{$U(4)\times U(2)_L\times U(2)_R\times USp(2)^2 $}\\
			\hline \hline			\rm{stack} & $N$ & $(n^1,l^1)\times(n^2,l^2)\times(n^3,l^3)$ & $n_{\Ysymm}$& $n_{\Yasymm}$ & $b$ & $b'$ & $c$ & $c'$ & 1 & 4\\
			\hline
			$a$ & 8 & $(-1,1)\times (-1,1)\times (-1,1)$ & 0 & 4  & 0 & 3 & -6 & 3 & 1 & 1\\
			$b$ & 4 & $(-2,1)\times (0,1)\times (-1,1)$ & -1 & 1  & - & - & -10 & 9 & 1 & 0\\
			$c$ & 4 & $(1,2)\times (-1,0)\times (-5,1)$ & -9 & 9  & - & - & - & - & 0 & 1\\
			\hline
			1 & 2 & $(1, 0)\times (1, 0)\times (2, 0)$& \multicolumn{8}{c|}{$x_A = 22x_B = 2x_C = \frac{11}{5}x_D$}\\
			4 & 2 & $(0, 1)\times (0, 1)\times (2, 0)$& \multicolumn{8}{c|}{$\beta^g_1=-3$, $\beta^g_4=-3$}\\
			& & & \multicolumn{8}{c|}{$\chi_1=\frac{1}{\sqrt{5}}$, $\chi_2=\frac{11}{\sqrt{5}}$, $\chi_3=4 \sqrt{5}$}\\
			\hline
		\end{tabular}
	\end{center}
\end{table}

\begin{table}\scriptsize
	\caption{D6-brane configurations and intersection numbers of Model 17, and its MSSM gauge coupling relation is $g^2_a=\frac{11}{6}g^2_b=\frac{5}{14}g^2_c=\frac{25}{52}(\frac{5}{3}g^2_Y)=\frac{8}{63} 5^{3/4} \sqrt{11} \pi  e^{\phi_4}$.}
	\label{tb:model17}
	\begin{center}
		\begin{tabular}{|c||c|c||c|c|c|c|c|c|c|c|}
			\hline\rm{Model} 17 & \multicolumn{10}{c|}{$U(4)\times U(2)_L\times U(2)_R\times USp(2)^2 $}\\
			\hline \hline			\rm{stack} & $N$ & $(n^1,l^1)\times(n^2,l^2)\times(n^3,l^3)$ & $n_{\Ysymm}$& $n_{\Yasymm}$ & $b$ & $b'$ & $c$ & $c'$ & 1 & 4\\
			\hline
			$a$ & 8 & $(-1,1)\times (-1,1)\times (-1,1)$ & 0 & 4  & 3 & 0 & -6 & 3 & 1 & 1\\
			$b$ & 4 & $(2,1)\times (0,1)\times (-1,-1)$ & 1 & -1  & - & - & -9 & 10 & 1 & 0\\
			$c$ & 4 & $(1,2)\times (-1,0)\times (-5,1)$ & -9 & 9  & - & - & - & - & 0 & 1\\
			\hline
			1 & 2 & $(1, 0)\times (1, 0)\times (2, 0)$& \multicolumn{8}{c|}{$x_A = 22x_B = 2x_C = \frac{11}{5}x_D$}\\
			4 & 2 & $(0, 1)\times (0, 1)\times (2, 0)$& \multicolumn{8}{c|}{$\beta^g_1=-3$, $\beta^g_4=-3$}\\
			& & & \multicolumn{8}{c|}{$\chi_1=\frac{1}{\sqrt{5}}$, $\chi_2=\frac{11}{\sqrt{5}}$, $\chi_3=4 \sqrt{5}$}\\
			\hline
		\end{tabular}
	\end{center}
\end{table}

\begin{table}\scriptsize
	\caption{D6-brane configurations and intersection numbers of Model 18, and its MSSM gauge coupling relation is $g^2_a=\frac{11}{6}g^2_b=\frac{5}{14}g^2_c=\frac{25}{52}(\frac{5}{3}g^2_Y)=\frac{8}{63} 5^{3/4} \sqrt{11} \pi  e^{\phi_4}$.}
	\label{tb:model18}
	\begin{center}
		\begin{tabular}{|c||c|c||c|c|c|c|c|c|c|c|}
			\hline\rm{Model} 18 & \multicolumn{10}{c|}{$U(4)\times U(2)_L\times U(2)_R\times USp(2)^2 $}\\
			\hline \hline			\rm{stack} & $N$ & $(n^1,l^1)\times(n^2,l^2)\times(n^3,l^3)$ & $n_{\Ysymm}$& $n_{\Yasymm}$ & $b$ & $b'$ & $c$ & $c'$ & 1 & 4\\
			\hline
			$a$ & 8 & $(-1,1)\times (-1,1)\times (-1,1)$ & 0 & 4  & 0 & 3 & 3 & -6 & 1 & 1\\
			$b$ & 4 & $(-2,1)\times (0,1)\times (-1,1)$ & -1 & 1  & - & - & 9 & -10 & 1 & 0\\
			$c$ & 4 & $(1,-2)\times (1,0)\times (5,1)$ & 9 & -9  & - & - & - & - & 0 & 1\\
			\hline
			1 & 2 & $(1, 0)\times (1, 0)\times (2, 0)$& \multicolumn{8}{c|}{$x_A = 22x_B = 2x_C = \frac{11}{5}x_D$}\\
			4 & 2 & $(0, 1)\times (0, 1)\times (2, 0)$& \multicolumn{8}{c|}{$\beta^g_1=-3$, $\beta^g_4=-3$}\\
			& & & \multicolumn{8}{c|}{$\chi_1=\frac{1}{\sqrt{5}}$, $\chi_2=\frac{11}{\sqrt{5}}$, $\chi_3=4 \sqrt{5}$}\\
			\hline
		\end{tabular}
	\end{center}
\end{table}

\begin{table}\scriptsize
	\caption{D6-brane configurations and intersection numbers of Model 19, and its MSSM gauge coupling relation is $g^2_a=\frac{11}{6}g^2_b=\frac{5}{14}g^2_c=\frac{25}{52}(\frac{5}{3}g^2_Y)=\frac{8}{63} 5^{3/4} \sqrt{11} \pi  e^{\phi_4}$.}
	\label{tb:model19}
	\begin{center}
		\begin{tabular}{|c||c|c||c|c|c|c|c|c|c|c|}
			\hline\rm{Model} 19 & \multicolumn{10}{c|}{$U(4)\times U(2)_L\times U(2)_R\times USp(2)^2 $}\\
			\hline \hline			\rm{stack} & $N$ & $(n^1,l^1)\times(n^2,l^2)\times(n^3,l^3)$ & $n_{\Ysymm}$& $n_{\Yasymm}$ & $b$ & $b'$ & $c$ & $c'$ & 1 & 4\\
			\hline
			$a$ & 8 & $(-1,1)\times (-1,1)\times (-1,1)$ & 0 & 4  & 3 & 0 & 3 & -6 & 1 & 1\\
			$b$ & 4 & $(2,1)\times (0,1)\times (-1,-1)$ & 1 & -1  & - & - & 10 & -9 & 1 & 0\\
			$c$ & 4 & $(1,-2)\times (1,0)\times (5,1)$ & 9 & -9  & - & - & - & - & 0 & 1\\
			\hline
			1 & 2 & $(1, 0)\times (1, 0)\times (2, 0)$& \multicolumn{8}{c|}{$x_A = 22x_B = 2x_C = \frac{11}{5}x_D$}\\
			4 & 2 & $(0, 1)\times (0, 1)\times (2, 0)$& \multicolumn{8}{c|}{$\beta^g_1=-3$, $\beta^g_4=-3$}\\
			& & & \multicolumn{8}{c|}{$\chi_1=\frac{1}{\sqrt{5}}$, $\chi_2=\frac{11}{\sqrt{5}}$, $\chi_3=4 \sqrt{5}$}\\
			\hline
		\end{tabular}
	\end{center}
\end{table}

\begin{table}\scriptsize
	\caption{D6-brane configurations and intersection numbers of Model 20, and its MSSM gauge coupling relation is $g^2_a=\frac{11}{6}g^2_b=\frac{5}{14}g^2_c=\frac{25}{52}(\frac{5}{3}g^2_Y)=\frac{8}{63} 5^{3/4} \sqrt{11} \pi  e^{\phi_4}$.}
	\label{tb:model20}
	\begin{center}
		\begin{tabular}{|c||c|c||c|c|c|c|c|c|c|c|}
			\hline\rm{Model} 20 & \multicolumn{10}{c|}{$U(4)\times U(2)_L\times U(2)_R\times USp(2)^2 $}\\
			\hline \hline			\rm{stack} & $N$ & $(n^1,l^1)\times(n^2,l^2)\times(n^3,l^3)$ & $n_{\Ysymm}$& $n_{\Yasymm}$ & $b$ & $b'$ & $c$ & $c'$ & 1 & 4\\
			\hline
			$a$ & 8 & $(-1,1)\times (-1,1)\times (-1,1)$ & 0 & 4  & 0 & 3 & -6 & 3 & 1 & 1\\
			$b$ & 4 & $(-2,1)\times (0,1)\times (-1,1)$ & -1 & 1  & - & - & -10 & 9 & 1 & 0\\
			$c$ & 4 & $(-1,-2)\times (1,0)\times (-5,1)$ & -9 & 9  & - & - & - & - & 0 & 1\\
			\hline
			1 & 2 & $(1, 0)\times (1, 0)\times (2, 0)$& \multicolumn{8}{c|}{$x_A = 22x_B = 2x_C = \frac{11}{5}x_D$}\\
			4 & 2 & $(0, 1)\times (0, 1)\times (2, 0)$& \multicolumn{8}{c|}{$\beta^g_1=-3$, $\beta^g_4=-3$}\\
			& & & \multicolumn{8}{c|}{$\chi_1=\frac{1}{\sqrt{5}}$, $\chi_2=\frac{11}{\sqrt{5}}$, $\chi_3=4 \sqrt{5}$}\\
			\hline
		\end{tabular}
	\end{center}
\end{table}

\begin{table}\scriptsize
	\caption{D6-brane configurations and intersection numbers of Model 21, and its MSSM gauge coupling relation is $g^2_a=\frac{11}{6}g^2_b=\frac{5}{14}g^2_c=\frac{25}{52}(\frac{5}{3}g^2_Y)=\frac{8}{63} 5^{3/4} \sqrt{11} \pi  e^{\phi_4}$.}
	\label{tb:model21}
	\begin{center}
		\begin{tabular}{|c||c|c||c|c|c|c|c|c|c|c|}
			\hline\rm{Model} 21 & \multicolumn{10}{c|}{$U(4)\times U(2)_L\times U(2)_R\times USp(2)^2 $}\\
			\hline \hline			\rm{stack} & $N$ & $(n^1,l^1)\times(n^2,l^2)\times(n^3,l^3)$ & $n_{\Ysymm}$& $n_{\Yasymm}$ & $b$ & $b'$ & $c$ & $c'$ & 1 & 4\\
			\hline
			$a$ & 8 & $(-1,1)\times (-1,1)\times (-1,1)$ & 0 & 4  & 3 & 0 & -6 & 3 & 1 & 1\\
			$b$ & 4 & $(2,1)\times (0,1)\times (-1,-1)$ & 1 & -1  & - & - & -9 & 10 & 1 & 0\\
			$c$ & 4 & $(-1,-2)\times (1,0)\times (-5,1)$ & -9 & 9  & - & - & - & - & 0 & 1\\
			\hline
			1 & 2 & $(1, 0)\times (1, 0)\times (2, 0)$& \multicolumn{8}{c|}{$x_A = 22x_B = 2x_C = \frac{11}{5}x_D$}\\
			4 & 2 & $(0, 1)\times (0, 1)\times (2, 0)$& \multicolumn{8}{c|}{$\beta^g_1=-3$, $\beta^g_4=-3$}\\
			& & & \multicolumn{8}{c|}{$\chi_1=\frac{1}{\sqrt{5}}$, $\chi_2=\frac{11}{\sqrt{5}}$, $\chi_3=4 \sqrt{5}$}\\
			\hline
		\end{tabular}
	\end{center}
\end{table}

\begin{table}\scriptsize
	\caption{D6-brane configurations and intersection numbers of Model 22, and its MSSM gauge coupling relation is $g^2_a=\frac{11}{6}g^2_b=\frac{5}{14}g^2_c=\frac{25}{52}(\frac{5}{3}g^2_Y)=\frac{8}{63} 5^{3/4} \sqrt{11} \pi  e^{\phi_4}$.}
	\label{tb:model22}
	\begin{center}
		\begin{tabular}{|c||c|c||c|c|c|c|c|c|c|c|}
			\hline\rm{Model} 22 & \multicolumn{10}{c|}{$U(4)\times U(2)_L\times U(2)_R\times USp(2)^2 $}\\
			\hline \hline			\rm{stack} & $N$ & $(n^1,l^1)\times(n^2,l^2)\times(n^3,l^3)$ & $n_{\Ysymm}$& $n_{\Yasymm}$ & $b$ & $b'$ & $c$ & $c'$ & 1 & 4\\
			\hline
			$a$ & 8 & $(-1,1)\times (-1,1)\times (-1,1)$ & 0 & 4  & 0 & 3 & -6 & 3 & 1 & 1\\
			$b$ & 4 & $(-2,1)\times (0,1)\times (-1,1)$ & -1 & 1  & - & - & -10 & 9 & 1 & 0\\
			$c$ & 4 & $(1,2)\times (1,0)\times (5,-1)$ & -9 & 9  & - & - & - & - & 0 & 1\\
			\hline
			1 & 2 & $(1, 0)\times (1, 0)\times (2, 0)$& \multicolumn{8}{c|}{$x_A = 22x_B = 2x_C = \frac{11}{5}x_D$}\\
			4 & 2 & $(0, 1)\times (0, 1)\times (2, 0)$& \multicolumn{8}{c|}{$\beta^g_1=-3$, $\beta^g_4=-3$}\\
			& & & \multicolumn{8}{c|}{$\chi_1=\frac{1}{\sqrt{5}}$, $\chi_2=\frac{11}{\sqrt{5}}$, $\chi_3=4 \sqrt{5}$}\\
			\hline
		\end{tabular}
	\end{center}
\end{table}

\begin{table}\scriptsize
	\caption{D6-brane configurations and intersection numbers of Model 23, and its MSSM gauge coupling relation is $g^2_a=\frac{11}{6}g^2_b=\frac{5}{14}g^2_c=\frac{25}{52}(\frac{5}{3}g^2_Y)=\frac{8}{63} 5^{3/4} \sqrt{11} \pi  e^{\phi_4}$.}
	\label{tb:model23}
	\begin{center}
		\begin{tabular}{|c||c|c||c|c|c|c|c|c|c|c|}
			\hline\rm{Model} 23 & \multicolumn{10}{c|}{$U(4)\times U(2)_L\times U(2)_R\times USp(2)^2 $}\\
			\hline \hline			\rm{stack} & $N$ & $(n^1,l^1)\times(n^2,l^2)\times(n^3,l^3)$ & $n_{\Ysymm}$& $n_{\Yasymm}$ & $b$ & $b'$ & $c$ & $c'$ & 1 & 4\\
			\hline
			$a$ & 8 & $(-1,1)\times (-1,1)\times (-1,1)$ & 0 & 4  & 3 & 0 & -6 & 3 & 1 & 1\\
			$b$ & 4 & $(2,1)\times (0,1)\times (-1,-1)$ & 1 & -1  & - & - & -9 & 10 & 1 & 0\\
			$c$ & 4 & $(1,2)\times (1,0)\times (5,-1)$ & -9 & 9  & - & - & - & - & 0 & 1\\
			\hline
			1 & 2 & $(1, 0)\times (1, 0)\times (2, 0)$& \multicolumn{8}{c|}{$x_A = 22x_B = 2x_C = \frac{11}{5}x_D$}\\
			4 & 2 & $(0, 1)\times (0, 1)\times (2, 0)$& \multicolumn{8}{c|}{$\beta^g_1=-3$, $\beta^g_4=-3$}\\
			& & & \multicolumn{8}{c|}{$\chi_1=\frac{1}{\sqrt{5}}$, $\chi_2=\frac{11}{\sqrt{5}}$, $\chi_3=4 \sqrt{5}$}\\
			\hline
		\end{tabular}
	\end{center}
\end{table}

\begin{table}\scriptsize
	\caption{D6-brane configurations and intersection numbers of Model 24, and its MSSM gauge coupling relation is $g^2_a=\frac{11}{6}g^2_b=\frac{5}{14}g^2_c=\frac{25}{52}(\frac{5}{3}g^2_Y)=\frac{8}{63} 5^{3/4} \sqrt{11} \pi  e^{\phi_4}$.}
	\label{tb:model24}
	\begin{center}
		\begin{tabular}{|c||c|c||c|c|c|c|c|c|c|c|}
			\hline\rm{Model} 24 & \multicolumn{10}{c|}{$U(4)\times U(2)_L\times U(2)_R\times USp(2)^2 $}\\
			\hline \hline			\rm{stack} & $N$ & $(n^1,l^1)\times(n^2,l^2)\times(n^3,l^3)$ & $n_{\Ysymm}$& $n_{\Yasymm}$ & $b$ & $b'$ & $c$ & $c'$ & 1 & 4\\
			\hline
			$a$ & 8 & $(-1,1)\times (-1,1)\times (-1,1)$ & 0 & 4  & 0 & 3 & 3 & -6 & 1 & 1\\
			$b$ & 4 & $(-2,1)\times (0,1)\times (-1,1)$ & -1 & 1  & - & - & 9 & -10 & 1 & 0\\
			$c$ & 4 & $(-1,2)\times (1,0)\times (-5,-1)$ & 9 & -9  & - & - & - & - & 0 & 1\\
			\hline
			1 & 2 & $(1, 0)\times (1, 0)\times (2, 0)$& \multicolumn{8}{c|}{$x_A = 22x_B = 2x_C = \frac{11}{5}x_D$}\\
			4 & 2 & $(0, 1)\times (0, 1)\times (2, 0)$& \multicolumn{8}{c|}{$\beta^g_1=-3$, $\beta^g_4=-3$}\\
			& & & \multicolumn{8}{c|}{$\chi_1=\frac{1}{\sqrt{5}}$, $\chi_2=\frac{11}{\sqrt{5}}$, $\chi_3=4 \sqrt{5}$}\\
			\hline
		\end{tabular}
	\end{center}
\end{table}

\begin{table}\scriptsize
	\caption{D6-brane configurations and intersection numbers of Model 25, and its MSSM gauge coupling relation is $g^2_a=3g^2_b=\frac{13}{5}g^2_c=\frac{65}{41}(\frac{5}{3}g^2_Y)=\frac{16}{5} \sqrt{3} \pi  e^{\phi_4}$.}
	\label{tb:model25}
	\begin{center}
		\begin{tabular}{|c||c|c||c|c|c|c|c|c|c|c|}
			\hline\rm{Model} 25 & \multicolumn{10}{c|}{$U(4)\times U(2)_L\times U(2)_R\times USp(2)\times USp(4) $}\\
			\hline \hline			\rm{stack} & $N$ & $(n^1,l^1)\times(n^2,l^2)\times(n^3,l^3)$ & $n_{\Ysymm}$& $n_{\Yasymm}$ & $b$ & $b'$ & $c$ & $c'$ & 1 & 3\\
			\hline
			$a$ & 8 & $(-1,1)\times (-1,0)\times (1,1)$ & 0 & 0  & 0 & 3 & -1 & -2 & 0 & 0\\
			$b$ & 4 & $(1,4)\times (0,1)\times (-1,-1)$ & -3 & 3  & - & - & -4 & 8 & 4 & 1\\
			$c$ & 4 & $(1,0)\times (1,-1)\times (1,3)$ & -2 & 2  & - & - & - & - & 0 & 1\\
			\hline
			1 & 2 & $(1, 0)\times (1, 0)\times (2, 0)$& \multicolumn{8}{c|}{$x_A = \frac{3}{4}x_B = \frac{1}{4}x_C = \frac{3}{4}x_D$}\\
			3 & 4 & $(0, 1)\times (1, 0)\times (0, 2)$& \multicolumn{8}{c|}{$\beta^g_1=-2$, $\beta^g_3=-4$}\\
			& & & \multicolumn{8}{c|}{$\chi_1=\frac{1}{2}$, $\chi_2=\frac{3}{2}$, $\chi_3=1$}\\
			\hline
		\end{tabular}
	\end{center}
\end{table}

\begin{table}\scriptsize
	\caption{D6-brane configurations and intersection numbers of Model 26, and its MSSM gauge coupling relation is $g^2_a=3g^2_b=\frac{13}{5}g^2_c=\frac{65}{41}(\frac{5}{3}g^2_Y)=\frac{16}{5} \sqrt{3} \pi  e^{\phi_4}$.}
	\label{tb:model26}
	\begin{center}
		\begin{tabular}{|c||c|c||c|c|c|c|c|c|c|c|}
			\hline\rm{Model} 26 & \multicolumn{10}{c|}{$U(4)\times U(2)_L\times U(2)_R\times USp(2)\times USp(4) $}\\
			\hline \hline			\rm{stack} & $N$ & $(n^1,l^1)\times(n^2,l^2)\times(n^3,l^3)$ & $n_{\Ysymm}$& $n_{\Yasymm}$ & $b$ & $b'$ & $c$ & $c'$ & 1 & 3\\
			\hline
			$a$ & 8 & $(-1,1)\times (-1,0)\times (1,1)$ & 0 & 0  & 3 & 0 & -2 & -1 & 0 & 0\\
			$b$ & 4 & $(-1,4)\times (0,1)\times (-1,1)$ & 3 & -3  & - & - & 4 & -8 & 4 & 1\\
			$c$ & 4 & $(1,0)\times (1,1)\times (1,-3)$ & 2 & -2  & - & - & - & - & 0 & 1\\
			\hline
			1 & 2 & $(1, 0)\times (1, 0)\times (2, 0)$& \multicolumn{8}{c|}{$x_A = \frac{3}{4}x_B = \frac{1}{4}x_C = \frac{3}{4}x_D$}\\
			3 & 4 & $(0, 1)\times (1, 0)\times (0, 2)$& \multicolumn{8}{c|}{$\beta^g_1=-2$, $\beta^g_3=-4$}\\
			& & & \multicolumn{8}{c|}{$\chi_1=\frac{1}{2}$, $\chi_2=\frac{3}{2}$, $\chi_3=1$}\\
			\hline
		\end{tabular}
	\end{center}
\end{table}

\begin{table}\scriptsize
	\caption{D6-brane configurations and intersection numbers of Model 27, and its MSSM gauge coupling relation is $g^2_a=3g^2_b=\frac{13}{5}g^2_c=\frac{65}{41}(\frac{5}{3}g^2_Y)=\frac{16}{5} \sqrt{3} \pi  e^{\phi_4}$.}
	\label{tb:model27}
	\begin{center}
		\begin{tabular}{|c||c|c||c|c|c|c|c|c|c|c|}
			\hline\rm{Model} 27 & \multicolumn{10}{c|}{$U(4)\times U(2)_L\times U(2)_R\times USp(2)\times USp(4) $}\\
			\hline \hline			\rm{stack} & $N$ & $(n^1,l^1)\times(n^2,l^2)\times(n^3,l^3)$ & $n_{\Ysymm}$& $n_{\Yasymm}$ & $b$ & $b'$ & $c$ & $c'$ & 1 & 3\\
			\hline
			$a$ & 8 & $(-1,1)\times (-1,0)\times (1,1)$ & 0 & 0  & 0 & 3 & -2 & -1 & 0 & 0\\
			$b$ & 4 & $(1,4)\times (0,1)\times (-1,-1)$ & -3 & 3  & - & - & 8 & -4 & 4 & 1\\
			$c$ & 4 & $(1,0)\times (1,1)\times (1,-3)$ & 2 & -2  & - & - & - & - & 0 & 1\\
			\hline
			1 & 2 & $(1, 0)\times (1, 0)\times (2, 0)$& \multicolumn{8}{c|}{$x_A = \frac{3}{4}x_B = \frac{1}{4}x_C = \frac{3}{4}x_D$}\\
			3 & 4 & $(0, 1)\times (1, 0)\times (0, 2)$& \multicolumn{8}{c|}{$\beta^g_1=-2$, $\beta^g_3=-4$}\\
			& & & \multicolumn{8}{c|}{$\chi_1=\frac{1}{2}$, $\chi_2=\frac{3}{2}$, $\chi_3=1$}\\
			\hline
		\end{tabular}
	\end{center}
\end{table}

\begin{table}\scriptsize
	\caption{D6-brane configurations and intersection numbers of Model 28, and its MSSM gauge coupling relation is $g^2_a=3g^2_b=\frac{13}{5}g^2_c=\frac{65}{41}(\frac{5}{3}g^2_Y)=\frac{16}{5} \sqrt{3} \pi  e^{\phi_4}$.}
	\label{tb:model28}
	\begin{center}
		\begin{tabular}{|c||c|c||c|c|c|c|c|c|c|c|}
			\hline\rm{Model} 28 & \multicolumn{10}{c|}{$U(4)\times U(2)_L\times U(2)_R\times USp(2)\times USp(4) $}\\
			\hline \hline			\rm{stack} & $N$ & $(n^1,l^1)\times(n^2,l^2)\times(n^3,l^3)$ & $n_{\Ysymm}$& $n_{\Yasymm}$ & $b$ & $b'$ & $c$ & $c'$ & 1 & 3\\
			\hline
			$a$ & 8 & $(-1,1)\times (-1,0)\times (1,1)$ & 0 & 0  & 3 & 0 & -1 & -2 & 0 & 0\\
			$b$ & 4 & $(-1,4)\times (0,1)\times (-1,1)$ & 3 & -3  & - & - & -8 & 4 & 4 & 1\\
			$c$ & 4 & $(-1,0)\times (-1,1)\times (1,3)$ & -2 & 2  & - & - & - & - & 0 & 1\\
			\hline
			1 & 2 & $(1, 0)\times (1, 0)\times (2, 0)$& \multicolumn{8}{c|}{$x_A = \frac{3}{4}x_B = \frac{1}{4}x_C = \frac{3}{4}x_D$}\\
			3 & 4 & $(0, 1)\times (1, 0)\times (0, 2)$& \multicolumn{8}{c|}{$\beta^g_1=-2$, $\beta^g_3=-4$}\\
			& & & \multicolumn{8}{c|}{$\chi_1=\frac{1}{2}$, $\chi_2=\frac{3}{2}$, $\chi_3=1$}\\
			\hline
		\end{tabular}
	\end{center}
\end{table}

\begin{table}\scriptsize
	\caption{D6-brane configurations and intersection numbers of Model 29, and its MSSM gauge coupling relation is $g^2_a=3g^2_b=\frac{13}{5}g^2_c=\frac{65}{41}(\frac{5}{3}g^2_Y)=\frac{16}{5} \sqrt{3} \pi  e^{\phi_4}$.}
	\label{tb:model29}
	\begin{center}
		\begin{tabular}{|c||c|c||c|c|c|c|c|c|c|c|}
			\hline\rm{Model} 29 & \multicolumn{10}{c|}{$U(4)\times U(2)_L\times U(2)_R\times USp(2)\times USp(4) $}\\
			\hline \hline			\rm{stack} & $N$ & $(n^1,l^1)\times(n^2,l^2)\times(n^3,l^3)$ & $n_{\Ysymm}$& $n_{\Yasymm}$ & $b$ & $b'$ & $c$ & $c'$ & 1 & 3\\
			\hline
			$a$ & 8 & $(-1,1)\times (-1,0)\times (1,1)$ & 0 & 0  & 0 & 3 & -1 & -2 & 0 & 0\\
			$b$ & 4 & $(1,4)\times (0,1)\times (-1,-1)$ & -3 & 3  & - & - & -4 & 8 & 4 & 1\\
			$c$ & 4 & $(-1,0)\times (-1,1)\times (1,3)$ & -2 & 2  & - & - & - & - & 0 & 1\\
			\hline
			1 & 2 & $(1, 0)\times (1, 0)\times (2, 0)$& \multicolumn{8}{c|}{$x_A = \frac{3}{4}x_B = \frac{1}{4}x_C = \frac{3}{4}x_D$}\\
			3 & 4 & $(0, 1)\times (1, 0)\times (0, 2)$& \multicolumn{8}{c|}{$\beta^g_1=-2$, $\beta^g_3=-4$}\\
			& & & \multicolumn{8}{c|}{$\chi_1=\frac{1}{2}$, $\chi_2=\frac{3}{2}$, $\chi_3=1$}\\
			\hline
		\end{tabular}
	\end{center}
\end{table}

\begin{table}\scriptsize
	\caption{D6-brane configurations and intersection numbers of Model 30, and its MSSM gauge coupling relation is $g^2_a=3g^2_b=\frac{13}{5}g^2_c=\frac{65}{41}(\frac{5}{3}g^2_Y)=\frac{16}{5} \sqrt{3} \pi  e^{\phi_4}$.}
	\label{tb:model30}
	\begin{center}
		\begin{tabular}{|c||c|c||c|c|c|c|c|c|c|c|}
			\hline\rm{Model} 30 & \multicolumn{10}{c|}{$U(4)\times U(2)_L\times U(2)_R\times USp(2)\times USp(4) $}\\
			\hline \hline			\rm{stack} & $N$ & $(n^1,l^1)\times(n^2,l^2)\times(n^3,l^3)$ & $n_{\Ysymm}$& $n_{\Yasymm}$ & $b$ & $b'$ & $c$ & $c'$ & 1 & 3\\
			\hline
			$a$ & 8 & $(-1,1)\times (-1,0)\times (1,1)$ & 0 & 0  & 3 & 0 & -2 & -1 & 0 & 0\\
			$b$ & 4 & $(-1,4)\times (0,1)\times (-1,1)$ & 3 & -3  & - & - & 4 & -8 & 4 & 1\\
			$c$ & 4 & $(-1,0)\times (-1,-1)\times (1,-3)$ & 2 & -2  & - & - & - & - & 0 & 1\\
			\hline
			1 & 2 & $(1, 0)\times (1, 0)\times (2, 0)$& \multicolumn{8}{c|}{$x_A = \frac{3}{4}x_B = \frac{1}{4}x_C = \frac{3}{4}x_D$}\\
			3 & 4 & $(0, 1)\times (1, 0)\times (0, 2)$& \multicolumn{8}{c|}{$\beta^g_1=-2$, $\beta^g_3=-4$}\\
			& & & \multicolumn{8}{c|}{$\chi_1=\frac{1}{2}$, $\chi_2=\frac{3}{2}$, $\chi_3=1$}\\
			\hline
		\end{tabular}
	\end{center}
\end{table}

\begin{table}\scriptsize
	\caption{D6-brane configurations and intersection numbers of Model 31, and its MSSM gauge coupling relation is $g^2_a=3g^2_b=\frac{13}{5}g^2_c=\frac{65}{41}(\frac{5}{3}g^2_Y)=\frac{16}{5} \sqrt{3} \pi  e^{\phi_4}$.}
	\label{tb:model31}
	\begin{center}
		\begin{tabular}{|c||c|c||c|c|c|c|c|c|c|c|}
			\hline\rm{Model} 31 & \multicolumn{10}{c|}{$U(4)\times U(2)_L\times U(2)_R\times USp(2)\times USp(4) $}\\
			\hline \hline			\rm{stack} & $N$ & $(n^1,l^1)\times(n^2,l^2)\times(n^3,l^3)$ & $n_{\Ysymm}$& $n_{\Yasymm}$ & $b$ & $b'$ & $c$ & $c'$ & 1 & 3\\
			\hline
			$a$ & 8 & $(-1,1)\times (-1,0)\times (1,1)$ & 0 & 0  & 0 & 3 & -2 & -1 & 0 & 0\\
			$b$ & 4 & $(1,4)\times (0,1)\times (-1,-1)$ & -3 & 3  & - & - & 8 & -4 & 4 & 1\\
			$c$ & 4 & $(-1,0)\times (-1,-1)\times (1,-3)$ & 2 & -2  & - & - & - & - & 0 & 1\\
			\hline
			1 & 2 & $(1, 0)\times (1, 0)\times (2, 0)$& \multicolumn{8}{c|}{$x_A = \frac{3}{4}x_B = \frac{1}{4}x_C = \frac{3}{4}x_D$}\\
			3 & 4 & $(0, 1)\times (1, 0)\times (0, 2)$& \multicolumn{8}{c|}{$\beta^g_1=-2$, $\beta^g_3=-4$}\\
			& & & \multicolumn{8}{c|}{$\chi_1=\frac{1}{2}$, $\chi_2=\frac{3}{2}$, $\chi_3=1$}\\
			\hline
		\end{tabular}
	\end{center}
\end{table}

\begin{table}\scriptsize
	\caption{D6-brane configurations and intersection numbers of Model 32, and its MSSM gauge coupling relation is $g^2_a=3g^2_b=\frac{13}{5}g^2_c=\frac{65}{41}(\frac{5}{3}g^2_Y)=\frac{16}{5} \sqrt{3} \pi  e^{\phi_4}$.}
	\label{tb:model32}
	\begin{center}
		\begin{tabular}{|c||c|c||c|c|c|c|c|c|c|c|}
			\hline\rm{Model} 32 & \multicolumn{10}{c|}{$U(4)\times U(2)_L\times U(2)_R\times USp(2)\times USp(4) $}\\
			\hline \hline			\rm{stack} & $N$ & $(n^1,l^1)\times(n^2,l^2)\times(n^3,l^3)$ & $n_{\Ysymm}$& $n_{\Yasymm}$ & $b$ & $b'$ & $c$ & $c'$ & 1 & 3\\
			\hline
			$a$ & 8 & $(-1,1)\times (-1,0)\times (1,1)$ & 0 & 0  & 3 & 0 & -2 & -1 & 0 & 0\\
			$b$ & 4 & $(-1,4)\times (0,1)\times (-1,1)$ & 3 & -3  & - & - & 4 & -8 & 4 & 1\\
			$c$ & 4 & $(-1,0)\times (1,1)\times (-1,3)$ & 2 & -2  & - & - & - & - & 0 & 1\\
			\hline
			1 & 2 & $(1, 0)\times (1, 0)\times (2, 0)$& \multicolumn{8}{c|}{$x_A = \frac{3}{4}x_B = \frac{1}{4}x_C = \frac{3}{4}x_D$}\\
			3 & 4 & $(0, 1)\times (1, 0)\times (0, 2)$& \multicolumn{8}{c|}{$\beta^g_1=-2$, $\beta^g_3=-4$}\\
			& & & \multicolumn{8}{c|}{$\chi_1=\frac{1}{2}$, $\chi_2=\frac{3}{2}$, $\chi_3=1$}\\
			\hline
		\end{tabular}
	\end{center}
\end{table}

\begin{table}\scriptsize
	\caption{D6-brane configurations and intersection numbers of Model 33, and its MSSM gauge coupling relation is $g^2_a=3g^2_b=\frac{13}{5}g^2_c=\frac{65}{41}(\frac{5}{3}g^2_Y)=\frac{16}{5} \sqrt{3} \pi  e^{\phi_4}$.}
	\label{tb:model33}
	\begin{center}
		\begin{tabular}{|c||c|c||c|c|c|c|c|c|c|c|}
			\hline\rm{Model} 33 & \multicolumn{10}{c|}{$U(4)\times U(2)_L\times U(2)_R\times USp(2)\times USp(4) $}\\
			\hline \hline			\rm{stack} & $N$ & $(n^1,l^1)\times(n^2,l^2)\times(n^3,l^3)$ & $n_{\Ysymm}$& $n_{\Yasymm}$ & $b$ & $b'$ & $c$ & $c'$ & 1 & 3\\
			\hline
			$a$ & 8 & $(-1,1)\times (-1,0)\times (1,1)$ & 0 & 0  & 0 & 3 & -2 & -1 & 0 & 0\\
			$b$ & 4 & $(1,4)\times (0,1)\times (-1,-1)$ & -3 & 3  & - & - & 8 & -4 & 4 & 1\\
			$c$ & 4 & $(-1,0)\times (1,1)\times (-1,3)$ & 2 & -2  & - & - & - & - & 0 & 1\\
			\hline
			1 & 2 & $(1, 0)\times (1, 0)\times (2, 0)$& \multicolumn{8}{c|}{$x_A = \frac{3}{4}x_B = \frac{1}{4}x_C = \frac{3}{4}x_D$}\\
			3 & 4 & $(0, 1)\times (1, 0)\times (0, 2)$& \multicolumn{8}{c|}{$\beta^g_1=-2$, $\beta^g_3=-4$}\\
			& & & \multicolumn{8}{c|}{$\chi_1=\frac{1}{2}$, $\chi_2=\frac{3}{2}$, $\chi_3=1$}\\
			\hline
		\end{tabular}
	\end{center}
\end{table}

\begin{table}\scriptsize
	\caption{D6-brane configurations and intersection numbers of Model 34, and its MSSM gauge coupling relation is $g^2_a=3g^2_b=\frac{13}{5}g^2_c=\frac{65}{41}(\frac{5}{3}g^2_Y)=\frac{16}{5} \sqrt{3} \pi  e^{\phi_4}$.}
	\label{tb:model34}
	\begin{center}
		\begin{tabular}{|c||c|c||c|c|c|c|c|c|c|c|}
			\hline\rm{Model} 34 & \multicolumn{10}{c|}{$U(4)\times U(2)_L\times U(2)_R\times USp(2)\times USp(4) $}\\
			\hline \hline			\rm{stack} & $N$ & $(n^1,l^1)\times(n^2,l^2)\times(n^3,l^3)$ & $n_{\Ysymm}$& $n_{\Yasymm}$ & $b$ & $b'$ & $c$ & $c'$ & 1 & 3\\
			\hline
			$a$ & 8 & $(-1,1)\times (-1,0)\times (1,1)$ & 0 & 0  & 3 & 0 & -1 & -2 & 0 & 0\\
			$b$ & 4 & $(-1,4)\times (0,1)\times (-1,1)$ & 3 & -3  & - & - & -8 & 4 & 4 & 1\\
			$c$ & 4 & $(-1,0)\times (1,-1)\times (-1,-3)$ & -2 & 2  & - & - & - & - & 0 & 1\\
			\hline
			1 & 2 & $(1, 0)\times (1, 0)\times (2, 0)$& \multicolumn{8}{c|}{$x_A = \frac{3}{4}x_B = \frac{1}{4}x_C = \frac{3}{4}x_D$}\\
			3 & 4 & $(0, 1)\times (1, 0)\times (0, 2)$& \multicolumn{8}{c|}{$\beta^g_1=-2$, $\beta^g_3=-4$}\\
			& & & \multicolumn{8}{c|}{$\chi_1=\frac{1}{2}$, $\chi_2=\frac{3}{2}$, $\chi_3=1$}\\
			\hline
		\end{tabular}
	\end{center}
\end{table}

\begin{table}\scriptsize
	\caption{D6-brane configurations and intersection numbers of Model 35, and its MSSM gauge coupling relation is $g^2_a=6g^2_b=\frac{26}{5}g^2_c=\frac{130}{67}(\frac{5}{3}g^2_Y)=\frac{16}{5} \sqrt{6} \pi  e^{\phi_4}$.}
	\label{tb:model35}
	\begin{center}
		\begin{tabular}{|c||c|c||c|c|c|c|c|c|c|}
			\hline\rm{Model} 35 & \multicolumn{9}{c|}{$U(4)\times U(2)_L\times U(2)_R\times USp(4) $}\\
			\hline \hline			\rm{stack} & $N$ & $(n^1,l^1)\times(n^2,l^2)\times(n^3,l^3)$ & $n_{\Ysymm}$& $n_{\Yasymm}$ & $b$ & $b'$ & $c$ & $c'$ & 3\\
			\hline
			$a$ & 8 & $(-1,1)\times (-1,0)\times (1,1)$ & 0 & 0  & 0 & 3 & -1 & -2 & 0\\
			$b$ & 4 & $(1,4)\times (0,1)\times (-1,-1)$ & -3 & 3  & - & - & -8 & 16 & 1\\
			$c$ & 4 & $(1,0)\times (2,-1)\times (1,3)$ & -5 & 5  & - & - & - & - & 1\\
			\hline
			3 & 4 & $(0, 1)\times (1, 0)\times (0, 2)$& \multicolumn{7}{c|}{$x_A = \frac{3}{2}x_B = \frac{1}{4}x_C = \frac{3}{2}x_D$}\\
			& & & \multicolumn{7}{c|}{$\beta^g_3=-4$}\\
			& & & \multicolumn{7}{c|}{$\chi_1=\frac{1}{2}$, $\chi_2=3$, $\chi_3=1$}\\
			\hline
		\end{tabular}
	\end{center}
\end{table}

\begin{table}\scriptsize
	\caption{D6-brane configurations and intersection numbers of Model 36, and its MSSM gauge coupling relation is $g^2_a=6g^2_b=\frac{26}{5}g^2_c=\frac{130}{67}(\frac{5}{3}g^2_Y)=\frac{16}{5} \sqrt{6} \pi  e^{\phi_4}$.}
	\label{tb:model36}
	\begin{center}
		\begin{tabular}{|c||c|c||c|c|c|c|c|c|c|}
			\hline\rm{Model} 36 & \multicolumn{9}{c|}{$U(4)\times U(2)_L\times U(2)_R\times USp(4) $}\\
			\hline \hline			\rm{stack} & $N$ & $(n^1,l^1)\times(n^2,l^2)\times(n^3,l^3)$ & $n_{\Ysymm}$& $n_{\Yasymm}$ & $b$ & $b'$ & $c$ & $c'$ & 3\\
			\hline
			$a$ & 8 & $(-1,1)\times (-1,0)\times (1,1)$ & 0 & 0  & 3 & 0 & -2 & -1 & 0\\
			$b$ & 4 & $(-1,4)\times (0,1)\times (-1,1)$ & 3 & -3  & - & - & 8 & -16 & 1\\
			$c$ & 4 & $(1,0)\times (2,1)\times (1,-3)$ & 5 & -5  & - & - & - & - & 1\\
			\hline
			3 & 4 & $(0, 1)\times (1, 0)\times (0, 2)$& \multicolumn{7}{c|}{$x_A = \frac{3}{2}x_B = \frac{1}{4}x_C = \frac{3}{2}x_D$}\\
			& & & \multicolumn{7}{c|}{$\beta^g_3=-4$}\\
			& & & \multicolumn{7}{c|}{$\chi_1=\frac{1}{2}$, $\chi_2=3$, $\chi_3=1$}\\
			\hline
		\end{tabular}
	\end{center}
\end{table}

\begin{table}\scriptsize
	\caption{D6-brane configurations and intersection numbers of Model 37, and its MSSM gauge coupling relation is $g^2_a=6g^2_b=\frac{26}{5}g^2_c=\frac{130}{67}(\frac{5}{3}g^2_Y)=\frac{16}{5} \sqrt{6} \pi  e^{\phi_4}$.}
	\label{tb:model37}
	\begin{center}
		\begin{tabular}{|c||c|c||c|c|c|c|c|c|c|}
			\hline\rm{Model} 37 & \multicolumn{9}{c|}{$U(4)\times U(2)_L\times U(2)_R\times USp(4) $}\\
			\hline \hline			\rm{stack} & $N$ & $(n^1,l^1)\times(n^2,l^2)\times(n^3,l^3)$ & $n_{\Ysymm}$& $n_{\Yasymm}$ & $b$ & $b'$ & $c$ & $c'$ & 3\\
			\hline
			$a$ & 8 & $(-1,1)\times (-1,0)\times (1,1)$ & 0 & 0  & 0 & 3 & -2 & -1 & 0\\
			$b$ & 4 & $(1,4)\times (0,1)\times (-1,-1)$ & -3 & 3  & - & - & 16 & -8 & 1\\
			$c$ & 4 & $(1,0)\times (2,1)\times (1,-3)$ & 5 & -5  & - & - & - & - & 1\\
			\hline
			3 & 4 & $(0, 1)\times (1, 0)\times (0, 2)$& \multicolumn{7}{c|}{$x_A = \frac{3}{2}x_B = \frac{1}{4}x_C = \frac{3}{2}x_D$}\\
			& & & \multicolumn{7}{c|}{$\beta^g_3=-4$}\\
			& & & \multicolumn{7}{c|}{$\chi_1=\frac{1}{2}$, $\chi_2=3$, $\chi_3=1$}\\
			\hline
		\end{tabular}
	\end{center}
\end{table}

\begin{table}\scriptsize
	\caption{D6-brane configurations and intersection numbers of Model 38, and its MSSM gauge coupling relation is $g^2_a=6g^2_b=\frac{26}{5}g^2_c=\frac{130}{67}(\frac{5}{3}g^2_Y)=\frac{16}{5} \sqrt{6} \pi  e^{\phi_4}$.}
	\label{tb:model38}
	\begin{center}
		\begin{tabular}{|c||c|c||c|c|c|c|c|c|c|}
			\hline\rm{Model} 38 & \multicolumn{9}{c|}{$U(4)\times U(2)_L\times U(2)_R\times USp(4) $}\\
			\hline \hline			\rm{stack} & $N$ & $(n^1,l^1)\times(n^2,l^2)\times(n^3,l^3)$ & $n_{\Ysymm}$& $n_{\Yasymm}$ & $b$ & $b'$ & $c$ & $c'$ & 3\\
			\hline
			$a$ & 8 & $(-1,1)\times (-1,0)\times (1,1)$ & 0 & 0  & 3 & 0 & -1 & -2 & 0\\
			$b$ & 4 & $(-1,4)\times (0,1)\times (-1,1)$ & 3 & -3  & - & - & -16 & 8 & 1\\
			$c$ & 4 & $(-1,0)\times (-2,1)\times (1,3)$ & -5 & 5  & - & - & - & - & 1\\
			\hline
			3 & 4 & $(0, 1)\times (1, 0)\times (0, 2)$& \multicolumn{7}{c|}{$x_A = \frac{3}{2}x_B = \frac{1}{4}x_C = \frac{3}{2}x_D$}\\
			& & & \multicolumn{7}{c|}{$\beta^g_3=-4$}\\
			& & & \multicolumn{7}{c|}{$\chi_1=\frac{1}{2}$, $\chi_2=3$, $\chi_3=1$}\\
			\hline
		\end{tabular}
	\end{center}
\end{table}

\begin{table}\scriptsize
	\caption{D6-brane configurations and intersection numbers of Model 39, and its MSSM gauge coupling relation is $g^2_a=6g^2_b=\frac{26}{5}g^2_c=\frac{130}{67}(\frac{5}{3}g^2_Y)=\frac{16}{5} \sqrt{6} \pi  e^{\phi_4}$.}
	\label{tb:model39}
	\begin{center}
		\begin{tabular}{|c||c|c||c|c|c|c|c|c|c|}
			\hline\rm{Model} 39 & \multicolumn{9}{c|}{$U(4)\times U(2)_L\times U(2)_R\times USp(4) $}\\
			\hline \hline			\rm{stack} & $N$ & $(n^1,l^1)\times(n^2,l^2)\times(n^3,l^3)$ & $n_{\Ysymm}$& $n_{\Yasymm}$ & $b$ & $b'$ & $c$ & $c'$ & 3\\
			\hline
			$a$ & 8 & $(-1,1)\times (-1,0)\times (1,1)$ & 0 & 0  & 0 & 3 & -1 & -2 & 0\\
			$b$ & 4 & $(1,4)\times (0,1)\times (-1,-1)$ & -3 & 3  & - & - & -8 & 16 & 1\\
			$c$ & 4 & $(-1,0)\times (-2,1)\times (1,3)$ & -5 & 5  & - & - & - & - & 1\\
			\hline
			3 & 4 & $(0, 1)\times (1, 0)\times (0, 2)$& \multicolumn{7}{c|}{$x_A = \frac{3}{2}x_B = \frac{1}{4}x_C = \frac{3}{2}x_D$}\\
			& & & \multicolumn{7}{c|}{$\beta^g_3=-4$}\\
			& & & \multicolumn{7}{c|}{$\chi_1=\frac{1}{2}$, $\chi_2=3$, $\chi_3=1$}\\
			\hline
		\end{tabular}
	\end{center}
\end{table}

\begin{table}\scriptsize
	\caption{D6-brane configurations and intersection numbers of Model 40, and its MSSM gauge coupling relation is $g^2_a=6g^2_b=\frac{26}{5}g^2_c=\frac{130}{67}(\frac{5}{3}g^2_Y)=\frac{16}{5} \sqrt{6} \pi  e^{\phi_4}$.}
	\label{tb:model40}
	\begin{center}
		\begin{tabular}{|c||c|c||c|c|c|c|c|c|c|}
			\hline\rm{Model} 40 & \multicolumn{9}{c|}{$U(4)\times U(2)_L\times U(2)_R\times USp(4) $}\\
			\hline \hline			\rm{stack} & $N$ & $(n^1,l^1)\times(n^2,l^2)\times(n^3,l^3)$ & $n_{\Ysymm}$& $n_{\Yasymm}$ & $b$ & $b'$ & $c$ & $c'$ & 3\\
			\hline
			$a$ & 8 & $(-1,1)\times (-1,0)\times (1,1)$ & 0 & 0  & 3 & 0 & -2 & -1 & 0\\
			$b$ & 4 & $(-1,4)\times (0,1)\times (-1,1)$ & 3 & -3  & - & - & 8 & -16 & 1\\
			$c$ & 4 & $(-1,0)\times (-2,-1)\times (1,-3)$ & 5 & -5  & - & - & - & - & 1\\
			\hline
			3 & 4 & $(0, 1)\times (1, 0)\times (0, 2)$& \multicolumn{7}{c|}{$x_A = \frac{3}{2}x_B = \frac{1}{4}x_C = \frac{3}{2}x_D$}\\
			& & & \multicolumn{7}{c|}{$\beta^g_3=-4$}\\
			& & & \multicolumn{7}{c|}{$\chi_1=\frac{1}{2}$, $\chi_2=3$, $\chi_3=1$}\\
			\hline
		\end{tabular}
	\end{center}
\end{table}

\begin{table}\scriptsize
	\caption{D6-brane configurations and intersection numbers of Model 41, and its MSSM gauge coupling relation is $g^2_a=6g^2_b=\frac{26}{5}g^2_c=\frac{130}{67}(\frac{5}{3}g^2_Y)=\frac{16}{5} \sqrt{6} \pi  e^{\phi_4}$.}
	\label{tb:model41}
	\begin{center}
		\begin{tabular}{|c||c|c||c|c|c|c|c|c|c|}
			\hline\rm{Model} 41 & \multicolumn{9}{c|}{$U(4)\times U(2)_L\times U(2)_R\times USp(4) $}\\
			\hline \hline			\rm{stack} & $N$ & $(n^1,l^1)\times(n^2,l^2)\times(n^3,l^3)$ & $n_{\Ysymm}$& $n_{\Yasymm}$ & $b$ & $b'$ & $c$ & $c'$ & 3\\
			\hline
			$a$ & 8 & $(-1,1)\times (-1,0)\times (1,1)$ & 0 & 0  & 0 & 3 & -2 & -1 & 0\\
			$b$ & 4 & $(1,4)\times (0,1)\times (-1,-1)$ & -3 & 3  & - & - & 16 & -8 & 1\\
			$c$ & 4 & $(-1,0)\times (-2,-1)\times (1,-3)$ & 5 & -5  & - & - & - & - & 1\\
			\hline
			3 & 4 & $(0, 1)\times (1, 0)\times (0, 2)$& \multicolumn{7}{c|}{$x_A = \frac{3}{2}x_B = \frac{1}{4}x_C = \frac{3}{2}x_D$}\\
			& & & \multicolumn{7}{c|}{$\beta^g_3=-4$}\\
			& & & \multicolumn{7}{c|}{$\chi_1=\frac{1}{2}$, $\chi_2=3$, $\chi_3=1$}\\
			\hline
		\end{tabular}
	\end{center}
\end{table}

\begin{table}\scriptsize
	\caption{D6-brane configurations and intersection numbers of Model 42, and its MSSM gauge coupling relation is $g^2_a=6g^2_b=\frac{26}{5}g^2_c=\frac{130}{67}(\frac{5}{3}g^2_Y)=\frac{16}{5} \sqrt{6} \pi  e^{\phi_4}$.}
	\label{tb:model42}
	\begin{center}
		\begin{tabular}{|c||c|c||c|c|c|c|c|c|c|}
			\hline\rm{Model} 42 & \multicolumn{9}{c|}{$U(4)\times U(2)_L\times U(2)_R\times USp(4) $}\\
			\hline \hline			\rm{stack} & $N$ & $(n^1,l^1)\times(n^2,l^2)\times(n^3,l^3)$ & $n_{\Ysymm}$& $n_{\Yasymm}$ & $b$ & $b'$ & $c$ & $c'$ & 3\\
			\hline
			$a$ & 8 & $(-1,1)\times (-1,0)\times (1,1)$ & 0 & 0  & 3 & 0 & -2 & -1 & 0\\
			$b$ & 4 & $(-1,4)\times (0,1)\times (-1,1)$ & 3 & -3  & - & - & 8 & -16 & 1\\
			$c$ & 4 & $(-1,0)\times (2,1)\times (-1,3)$ & 5 & -5  & - & - & - & - & 1\\
			\hline
			3 & 4 & $(0, 1)\times (1, 0)\times (0, 2)$& \multicolumn{7}{c|}{$x_A = \frac{3}{2}x_B = \frac{1}{4}x_C = \frac{3}{2}x_D$}\\
			& & & \multicolumn{7}{c|}{$\beta^g_3=-4$}\\
			& & & \multicolumn{7}{c|}{$\chi_1=\frac{1}{2}$, $\chi_2=3$, $\chi_3=1$}\\
			\hline
		\end{tabular}
	\end{center}
\end{table}

\begin{table}\scriptsize
	\caption{D6-brane configurations and intersection numbers of Model 43, and its MSSM gauge coupling relation is $g^2_a=6g^2_b=\frac{26}{5}g^2_c=\frac{130}{67}(\frac{5}{3}g^2_Y)=\frac{16}{5} \sqrt{6} \pi  e^{\phi_4}$.}
	\label{tb:model43}
	\begin{center}
		\begin{tabular}{|c||c|c||c|c|c|c|c|c|c|}
			\hline\rm{Model} 43 & \multicolumn{9}{c|}{$U(4)\times U(2)_L\times U(2)_R\times USp(4) $}\\
			\hline \hline			\rm{stack} & $N$ & $(n^1,l^1)\times(n^2,l^2)\times(n^3,l^3)$ & $n_{\Ysymm}$& $n_{\Yasymm}$ & $b$ & $b'$ & $c$ & $c'$ & 3\\
			\hline
			$a$ & 8 & $(-1,1)\times (-1,0)\times (1,1)$ & 0 & 0  & 0 & 3 & -2 & -1 & 0\\
			$b$ & 4 & $(1,4)\times (0,1)\times (-1,-1)$ & -3 & 3  & - & - & 16 & -8 & 1\\
			$c$ & 4 & $(-1,0)\times (2,1)\times (-1,3)$ & 5 & -5  & - & - & - & - & 1\\
			\hline
			3 & 4 & $(0, 1)\times (1, 0)\times (0, 2)$& \multicolumn{7}{c|}{$x_A = \frac{3}{2}x_B = \frac{1}{4}x_C = \frac{3}{2}x_D$}\\
			& & & \multicolumn{7}{c|}{$\beta^g_3=-4$}\\
			& & & \multicolumn{7}{c|}{$\chi_1=\frac{1}{2}$, $\chi_2=3$, $\chi_3=1$}\\
			\hline
		\end{tabular}
	\end{center}
\end{table}

\begin{table}\scriptsize
	\caption{D6-brane configurations and intersection numbers of Model 44, and its MSSM gauge coupling relation is $g^2_a=6g^2_b=\frac{26}{5}g^2_c=\frac{130}{67}(\frac{5}{3}g^2_Y)=\frac{16}{5} \sqrt{6} \pi  e^{\phi_4}$.}
	\label{tb:model44}
	\begin{center}
		\begin{tabular}{|c||c|c||c|c|c|c|c|c|c|}
			\hline\rm{Model} 44 & \multicolumn{9}{c|}{$U(4)\times U(2)_L\times U(2)_R\times USp(4) $}\\
			\hline \hline			\rm{stack} & $N$ & $(n^1,l^1)\times(n^2,l^2)\times(n^3,l^3)$ & $n_{\Ysymm}$& $n_{\Yasymm}$ & $b$ & $b'$ & $c$ & $c'$ & 3\\
			\hline
			$a$ & 8 & $(-1,1)\times (-1,0)\times (1,1)$ & 0 & 0  & 3 & 0 & -1 & -2 & 0\\
			$b$ & 4 & $(-1,4)\times (0,1)\times (-1,1)$ & 3 & -3  & - & - & -16 & 8 & 1\\
			$c$ & 4 & $(-1,0)\times (2,-1)\times (-1,-3)$ & -5 & 5  & - & - & - & - & 1\\
			\hline
			3 & 4 & $(0, 1)\times (1, 0)\times (0, 2)$& \multicolumn{7}{c|}{$x_A = \frac{3}{2}x_B = \frac{1}{4}x_C = \frac{3}{2}x_D$}\\
			& & & \multicolumn{7}{c|}{$\beta^g_3=-4$}\\
			& & & \multicolumn{7}{c|}{$\chi_1=\frac{1}{2}$, $\chi_2=3$, $\chi_3=1$}\\
			\hline
		\end{tabular}
	\end{center}
\end{table}

\end{document}